\documentclass[a4paper,fleqn,usenatbib]{mnras}

\usepackage{newtxtext,newtxmath}

\usepackage[T1]{fontenc}
\usepackage{ae,aecompl}
\usepackage{subfigure}

\usepackage{graphicx}	
\usepackage{amsmath}	
\usepackage{amssymb}	

\title[Primordial Structure: Orion A vs. Pipe]{Not so different after all: Properties and Spatial Structure of Column Density Peaks in the Pipe and Orion A Clouds}

\author[C. G. Rom\'an-Z\'u\~niga et al.]{
Carlos G. Rom\'an-Z\'u\~niga,$^{1}$\thanks{E-mail: croman@astro.unam.mx}
Emilio Alfaro,$^{2}$
Aina Palau,$^{3}$
Birgit Hasenberger,$^{4}$
\newauthor 
Jo\~ao F. Alves,$^{4}$ 
Marco Lombardi,$^{5}$ 
and G. Paloma S. S\'anchez$^{3}$
\\
$^{1}$Instituto de Astronom\'ia en Ensenada, Universidad Nacional Aut\'onoma de M\'exico, \\
Km 107 Carr. Tijuana-Ensenada, Ensenada BC 22860, Mexico\\
$^{2}$Instituto de Astrof\'isica de Andaluc\'ia, (CSIC), Granada, 18008, Spain\\
$^{3}$Instituto de Radioastronom\'ia y Astrof\'isica, UNAM, Morelia, Mexico\\
$^{4}$Department for Astrophysics, University of Vienna, Vienna 1180, Austria\\
$^{5}$Dipartimento di Fisica, Universit\`a di Milano, Milan, Italy\\
}

\date{Accepted XXX. Received YYY; in original form ZZZ}

\pubyear{2019}

\begin{document}
\label{firstpage}
\pagerange{\pageref{firstpage}--\pageref{lastpage}}
\maketitle

\begin{abstract}

We present a comparative study of the physical properties and the spatial distribution of column density peaks in two Giant Molecular Clouds (GMC), the Pipe Nebula and Orion A, which exemplify opposite cases of star cluster formation stages. The density peaks were extracted from dust extinction maps constructed from Herschel/SPIRE farinfrared images. We compare the distribution functions for dust temperature, mass, equivalent radius and mean volume density of peaks in both clouds, and made a more fair comparison by isolating the less active Tail region in Orion A and by convolving the Pipe Nebula map to simulate placing it at a distance similar to that of the Orion Complex. The peak mass distributions for Orion A, the Tail, and the convolved Pipe, have similar ranges, sharing a maximum near 5 M$_\odot$, and a similar power law drop above 10 M$_\odot$. Despite the clearly distinct evolutive stage of the clouds, there are very important similarities in the physical and spatial distribution properties of the column density peaks, pointing to a scenario where they form as a result of uniform fragmentation of filamentary structures across the various scales of the cloud, with density being the parameter leading the fragmentation, and with clustering being a direct result of thermal fragmentation at different spatial scales. Our work strongly supports the idea that the formation of clusters in GMC could be the result of the primordial organization of pre-stellar material.

\end{abstract}

\begin{keywords}
stars:formation -- infrared:ISM -- ISM:structure -- ISM:clouds
\end{keywords}



\section{Introduction} \label{sec:intro}

In the currently accepted model of star formation, most stars form in families of star clusters within Giant Molecular Clouds (GMCs). However, 
clusters and their families can be quite diverse, and there is no consensus yet on a general model for cluster formation. An additional complication is that once stars are born, they destroy most of the evidence relating to the initial stages of their formation. In this sense, determining the early stages of cluster formation in GMCs may be key to understanding many distinct and complex processes involved in such a phenomenon. The Pipe Nebula, a nearby (d=130 pc) GMC with a total mass of $7.9\times10^3\mathrm{\ M_\odot}$ \citep{lada:2010aa} and very little star formation activity \citep[][]{brooke:2007aa,forbrich:2009aa,forbrich:2010aa} has played a workhorse role in a number of previous studies \citep[e.g.][]{alves:2007aa,muench:2007aa,lada:2008aa,falves:2008aa,rathborne:2009aa,roman:2009aa,roman:2010aa,peretto:2012aa,frau:2015aa,birgit2018}. Many of these works have the common goal of determining how GMCs are organized in the stages that precede their collapse and the consequent conversion of gas into stars. Interestingly, despite the fact that it is almost starless, the column density peaks\footnote{We avoid here to use the word ``core", to be consistent with the prescription of \citet{rathborne:2009aa}, that requires individual cores to be defined in terms of a separation of at least one Jeans length and a velocity difference equal or larger to the sound speed. Also, due to the limited resolution, we cannot resolve every molecular clump into individual cores. The use of the more general ``density peak" nomenclature allows us to include significant prestellar and protostellar structures where these requirements have not been confirmed in advance to our analysis.} in the Pipe appear to show a clear imprint of clustering, as well as segregation by density and mass \citep[e.g.][hereafter RAL10 and AR18, respectively]{roman:2010aa,alfaro:2018aa}. The Pipe was also the first cloud where a ``Core Mass Function" \citep[CMF, see][]{alves:2007aa,rathborne:2009aa} was constructed, leading to many subsequent studies that have discussed what appears to be a clear homology to its stellar counterpart \citep[e.g.][]{alves:2007aa,goodwin:2008aa,swift:2008aa,kainu:2009aa}. These findings are suggestive of a primordial organization of clouds towards the formation of groups and clusters, evident in the properties of density peaks. 

The low star forming activity in the Pipe appears, however, to be a rather unusual case among nearby GMCs. Star formation is plentiful in the solar vicinity, GMCs in the Gould Belt have formed clusters and, even the Aquila and Lupus clouds, once considered as bookcase examples of GMCs with low activity formation and starless regions \citep[][]{tachihara:2001,prato:2008aa,benedettini:2015aa}, have significantly more star formation activity than the Pipe. It has been suggested that this cloud is chemically evolved \citep{frau:2012aa} with some of its dense cores undergoing collapse \citep{juarez:2017aa}, but some processes (e.g. a magnetic field) could be retarding the collapse of the cloud \citep[][]{falves:2008aa,franco:2010,roman:2012,bailey:2012,frau:2015aa,li:2015}. It becomes evident that we have to examine the Pipe in the context of other more evolved clouds, in order to have more clues as to whether primordial clustering properties are common among distinct GMCs.

\begin{figure*}
\includegraphics[width=0.99\textwidth, angle=0]{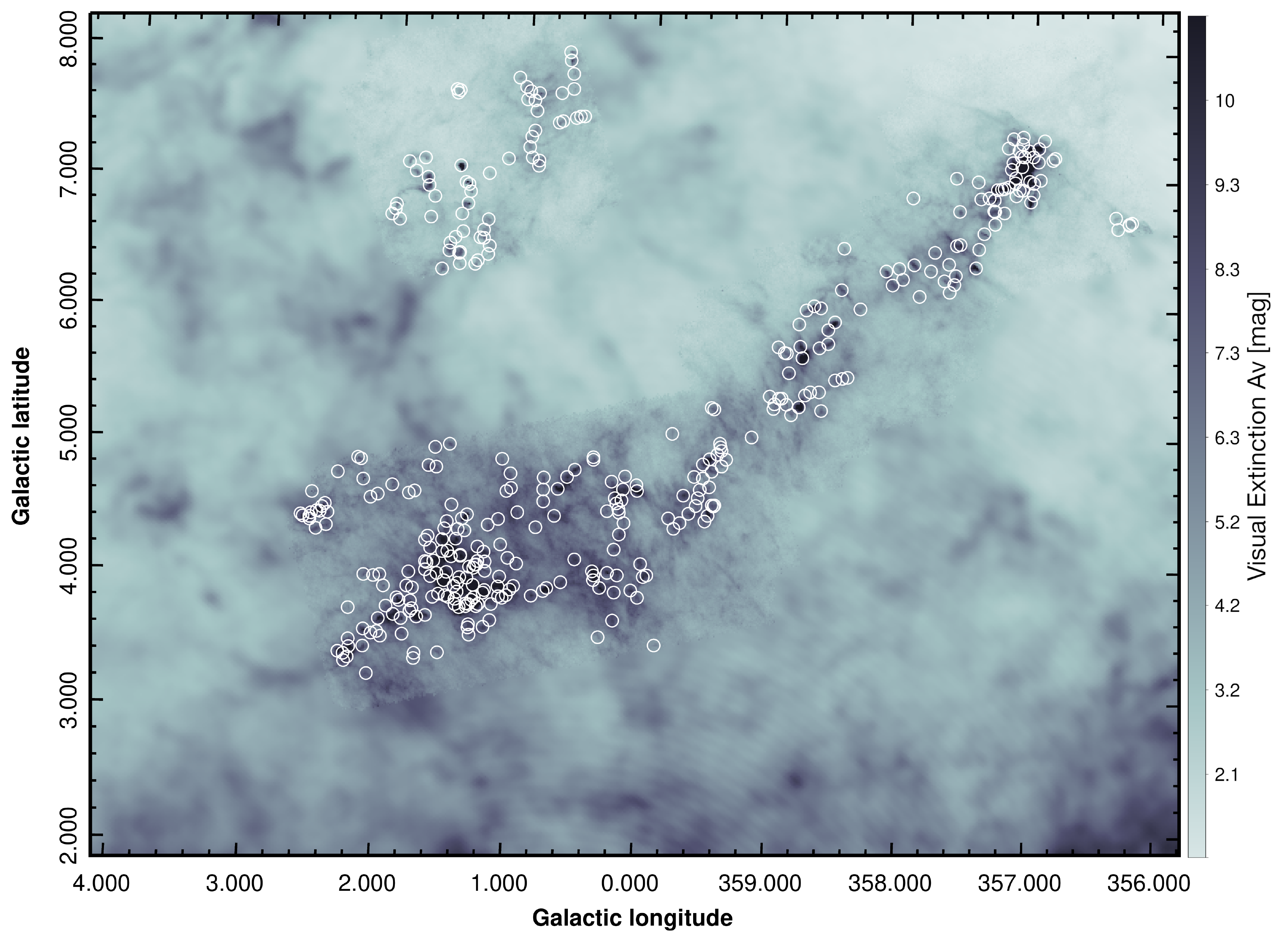}
\caption{A dust extinction map of the Pipe Nebula molecular cloud, derived from Herschel-SPIRE/Planck/2MASS infrared maps. The colour scale indicates visual extinction $A_V$ in magnitudes. It is possible to notice the coverage by Herschel, stitched onto the lower resolution background derived from Planck dust emission maps. The white circles indicate the locations of density peaks identified with our technique.
\label{fig:PN_Av}}
\end{figure*}

\par
\begin{figure*}
\includegraphics[width=1.0\textwidth, angle=0]{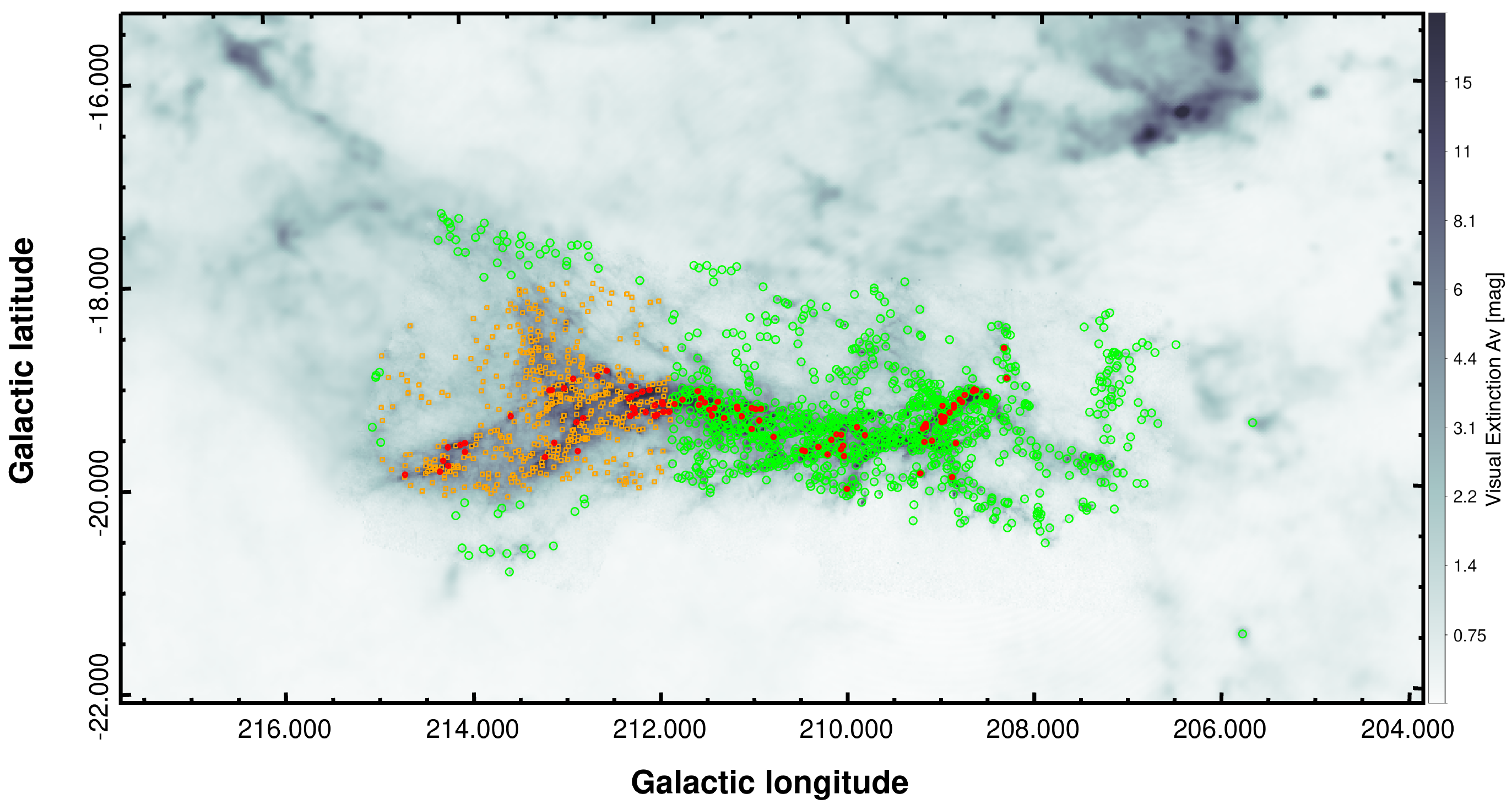}
\caption{A dust extinction map of the Orion A molecular cloud derived from Herschel-SPIRE/Planck/2MASS infrared maps. 
The symbols indicate the locations of density peaks identified with our technique; orange squares indicate the locations of density peaks the Tail sub-region, green circles are peaks in the rest of the cloud, and red dots indicate positions of peaks coincident with YSOs from the list of \citet{megeath:2012aa}. The feature near $(l,b)=(206,-15)$ is part of the Orion B complex, at the lower resolution of the Planck dust emission map that conforms the background.\label{fig:OA_Av}}
\end{figure*}

 One interesting place to make a comparison is the Orion A molecular Cloud, the well known star cluster forming complex located at d=380 pc, with a total mass of $6.8\times10^4\mathrm{\ M_\odot}$ \citep{lada:2010aa}. In principle, the Pipe and the Orion A may look as \textit{opposite} cases in terms of their star forming activity level, size, mass and the presence of massive sources. However, they also have some important similarities. For instance, the Orion A GMC presents, just as the Pipe, a clear filamentary structure (see Figures \ref{fig:PN_Av} and \ref{fig:OA_Av}). Moreover, the Orion A appears to be divided in two sections: its west end hosts a massive star cluster \citep[the copiously studied Orion Nebula Cluster or ONC:\ ][\ just to cite a few examples]{hillenbrand:1998,garmire:2000,huff:2006,morales:2011,kounkel:2018aa,pavlik19}, while its east end is much less active, with only a few young star groups \citep[see\ ][\ and references therein]{meingast:2016aa}. The Pipe also contains its only star cluster, B59, at its Western end. Perhaps the Orion A Complex is a transition case between a low-active cloud like the Pipe and more active GMCs --like those forming massive star clusters--, and we may find evidence of the transition from primordial structure to active star cluster formation by comparing both clouds.

In this paper we apply a simple, but consistent analysis of peaks identified and characterised from 2-dimensional column density and dust temperature maps of the Pipe and Orion A GMCs, constructed by means of a combination of Herschel\footnote{Herschel is an ESA space observatory with science instruments provided by European-led Principal Investigator consortia and with important participation from NASA} dust emission, Planck dust emission and 2MASS dust extinction maps using the methods of \citet{lombardi:2014aa} and \citet{birgit2018} (for Orion A and the Pipe maps, respectively). These Herschel-Planck-2MASS (hereafter HP2) maps provide a well calibrated estimation of total opacity across large areas.  We use these HP2 maps to make a comparison, as direct as possible, of the physical and spatial distribution properties of density peaks in both the Pipe and Orion A GMCs, and how such distributions are related to the primordial organisation of clouds toward the formation of star clusters.

The paper is divided as follows: after this brief introduction, we describe our data and methods in section \ref{s:data}, followed by a presentation of results in section \ref{s:results}. Finally, we discuss our findings in section \ref{s:discussion}.

\section{Data Processing}\label{s:data}

The HP2 maps were processed into FITS files containing two extension layers, stored as individual images with an angular resolution (FWHM) of 36 arcsec (corresponding to Herschel/SPIRE 500$\mu$ band) and a pixel resolution of $15\arcsec$. The first layer is a 850 $\mu$m optical depth $\tau _{850}$ map, and the second is a colour temperature, $T_d$ map. We show the dust extinction maps derived from $\tau _{850}$, in Figures \ref{fig:PN_Av} and \ref{fig:OA_Av}. As described in \citet{lombardi:2014aa} and \citet{birgit2018}, it is possible to convert from opacity to extinction using a direct linear conversion $A_K=\gamma\tau _{850}+\delta$ with respect to a near-infrared excess extinction 
\citep[NICEST;][]{lombardi:2009aa} map . The ordinate $\delta$ is related to uncertainties in the flux calibration of Herschel or/and in the intrinsic colours for the NICEST map. In our case, these uncertainties were reconsidered as small modifications in the value $\gamma$.  The final linear slope, $\gamma\approx 1.0857C_{2.2}/\kappa_{850}$ we use, relates both $C_{2.2}$, the extinction coefficient at 2.2 $\mu$m and $\kappa_{850}$, the opacity at 850 $\mu$m. In our case, $\gamma$ has a value of 2637.06 mag for the Orion A map, and 4756.27 mag for the Pipe map. We chose to use visual extinction, $A_V$, to make our analysis compatible with previous studies (e.g. RAL10), using a conversion coefficient of $(A_K/A_V)=0.112$ from the extinction law of \citet{RL85}. The resultant $A_V$ maps were processed following the prescription by RAL10, which we briefly summarise below:

\subsection{Wavelet filtering}\label{s:data:ss:mvm}

 Each $A_V$ map was processed with a code that makes use of the Multi Scale Vision Model algorithm (MVM) of \citet{rb:1997}. This code (Benoit Vandame, personal communication) uses a wavelet filter to remove large scale emission of low significance and to highlight and isolate significant regions in a 2D map. In our case, we are interested in highlighting peaks and the filaments on which they are located. The algorithm is trained to focus on roundish, coherent structures standing over a larger scale, lower density background. The significant regions are, in our case, reconstructed with the MVM algorithm using a 10-$\sigma$ significance tree at four pixel scales, $2^{l} \times s_p$, with $l={1,..,4}$ and $s_p$ being the pixel scale resolution of the maps ($15\arcsec$). The filtering is done in the pixel scale, so that the physical scales removed in Orion A and the Pipe Nebula are not the same, but careful examination of the resultant images confirmed that the selected parameter values for the  MVM algorithm kept in both cases, similar levels on the filamentary structures that in turn contain the density peaks, as we show in appendix \ref{app}.  

\subsection{Peak identification}\label{s:data:ss:clf2d}
We then processed the filtered images with the \texttt{CLUMPFIND-2D} algorithm of \citet{williams:1994aa}, which can identify individual peaks as local maxima and separates adjacent regions by defining their individual boundaries using contour levels. In our case, we made contour levels at constant steps of 10$\sigma_{A_V}$ within $0<(A_V/\mathrm{mag})<30$, where $\sigma_{A_V}$ was estimated as the standard deviation of pixel values within a small (5$\arcmin$) circular aperture in a region of each map where we could not distinguish any pixel groupings above the low noise background emission of the HP2 maps. Additionally, in order to avoid spurious detections, we set the algorithm to define a peak as a region with a minimum area of 30 pixels. The \texttt{CLUMPFIND-2D} algorithm has proven to be reliable and to provide satisfactory results when used on MVM processed extinction maps in previous studies \citep[e.g.][, RAL10]{rathborne:2009aa,roman:2009aa}, but more importantly, it gives us a simple, consistent methodology with a direct way to estimate some physical properties: each peak in the list is provided with a total $A_V$ flux, an equivalent radius and a position. The code also produces a map where each peak area is separated into regions. From these, we estimated mass and mean volume density using a standard conversion from $A_V$ to $N_H$ to mass (see equation 2 in RAL10) and converted equivalent radii to units of pc using a fixed distance (in our case 130 and 380 pc for the Pipe and Orion A clouds, respectively); we also estimated the mean colour temperature within each peak boundary. The processing yields 371 and 1573 density peak identifications for the Pipe and the Orion A $A_V$ maps, respectively.

\begin{figure*}
\includegraphics[width=3.5in, angle=0]{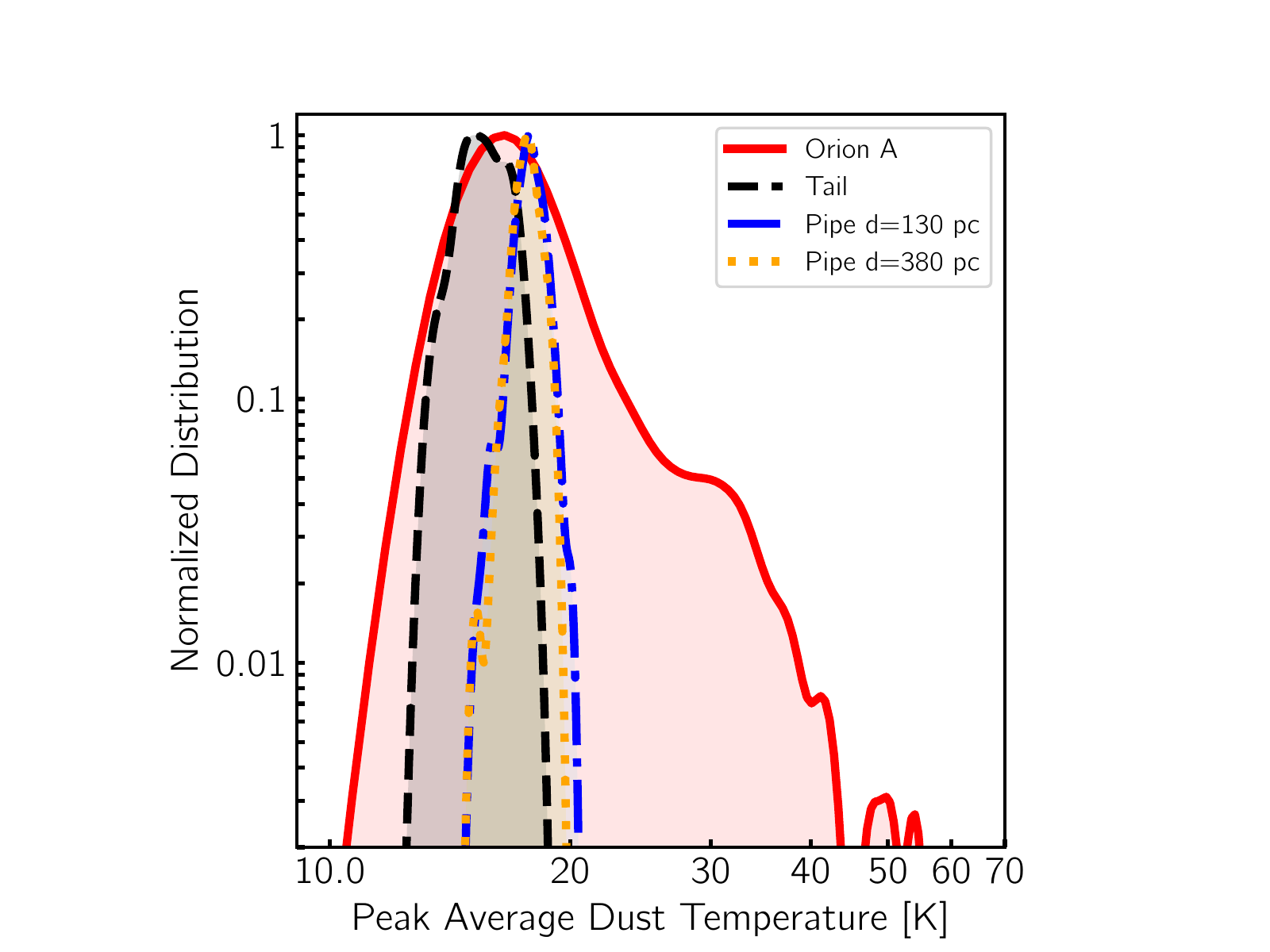}\includegraphics[width=3.51in, angle=0]{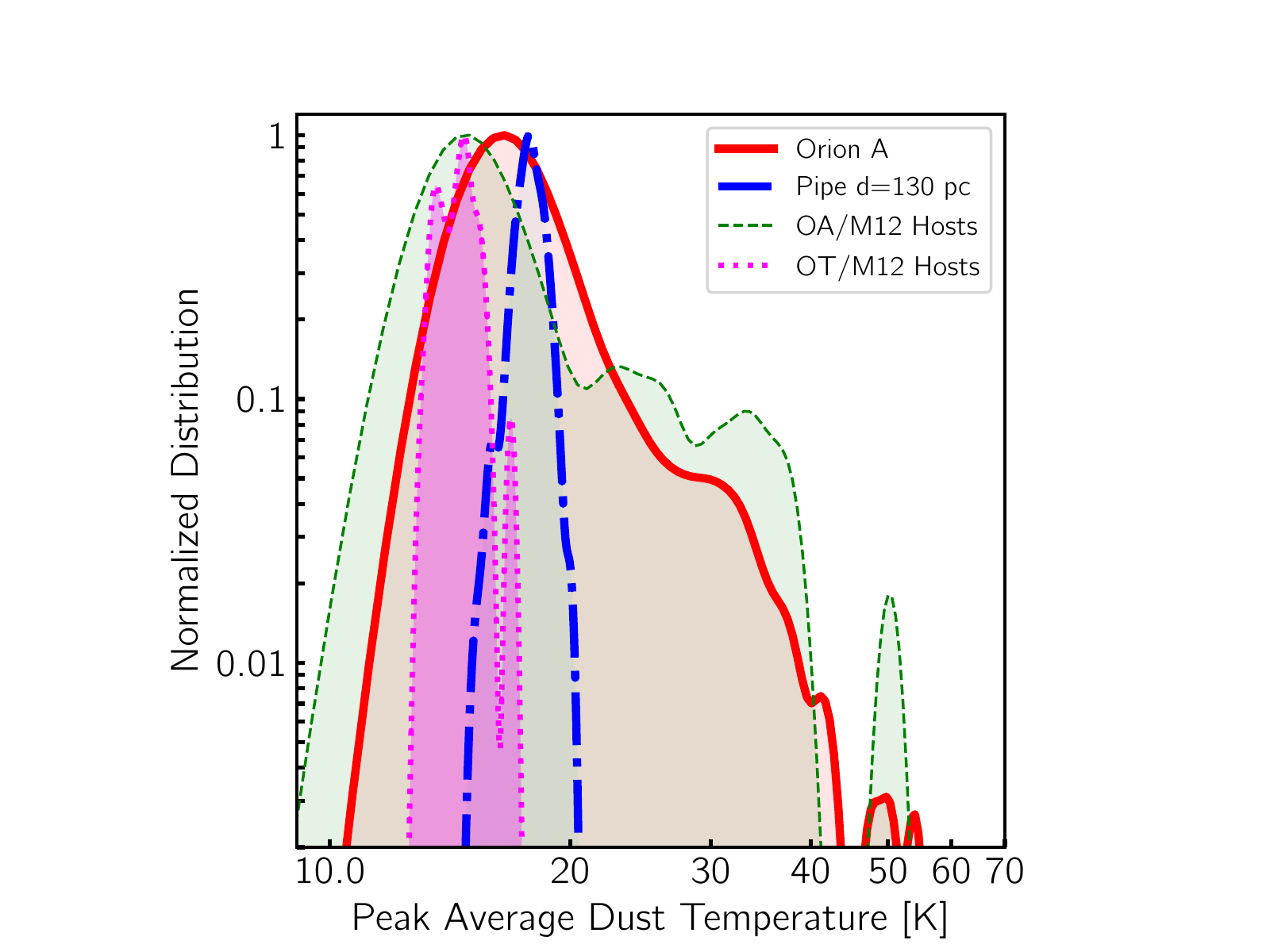}
\caption{Distribution of average Dust Temperature for the Pipe and Orion A column density peaks. \textit{Left}: The blue KDE corresponds to the Pipe at d=130 pc; the red KDE corresponds to the whole Orion A cloud at a distance of d=380 pc; the black KDE is for the Orion A Tail region; the yellow KDE is for the Pipe Nebula after convolving it to simulate placing it at the distance of Orion A. \textit{Right}: the magenta and light green KDEs corresponds to peaks hosting YSOs from the list of \citet{megeath:2012aa} in the Orion A and OATail, and we show, for comparison, the KDE for the Pipe and Orion A \label{fig:TdustHist}}
\end{figure*}

\begin{figure*}
\includegraphics[width=3.2in, angle=0]{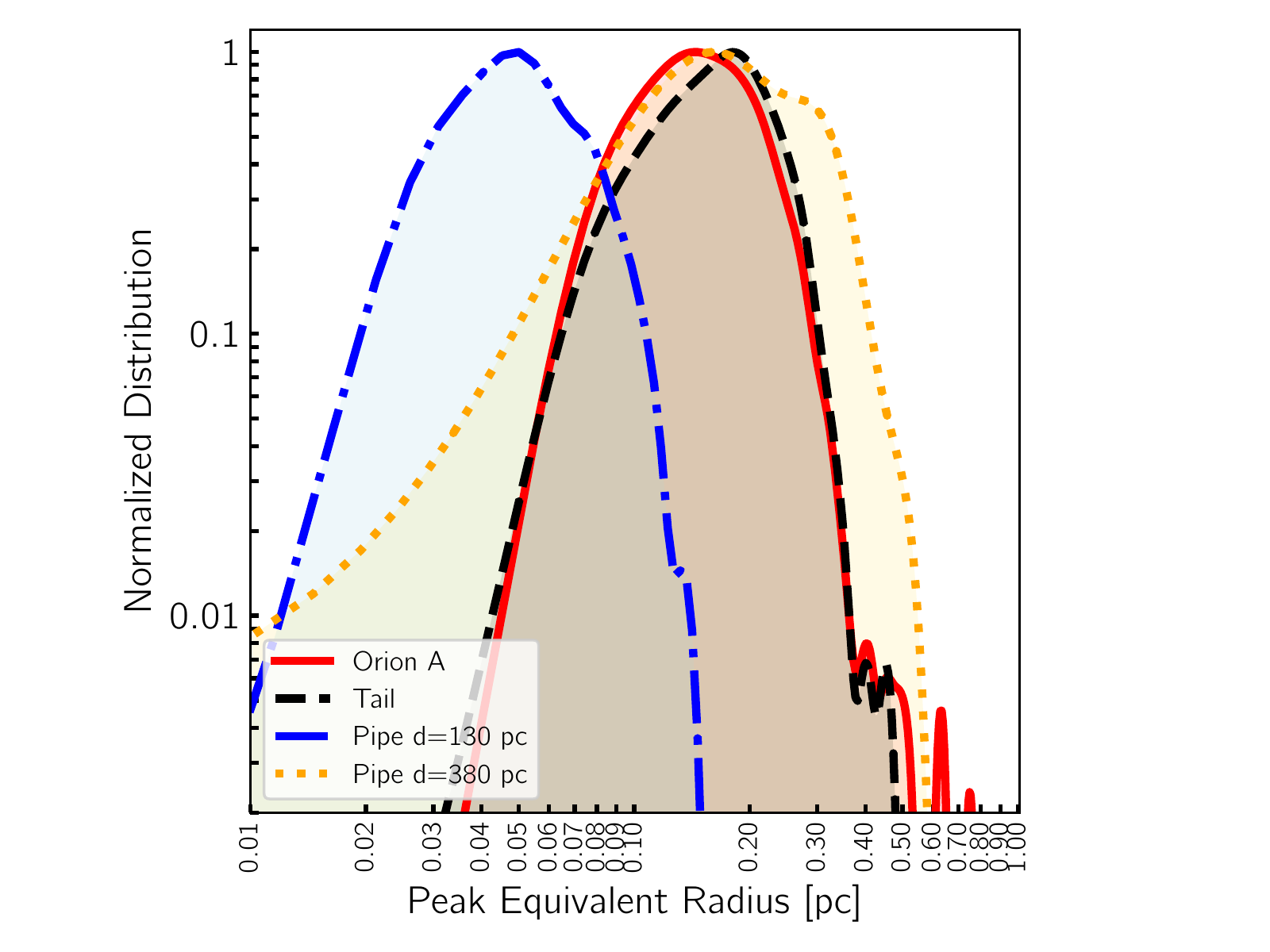}\includegraphics[width=3.18in, angle=0]{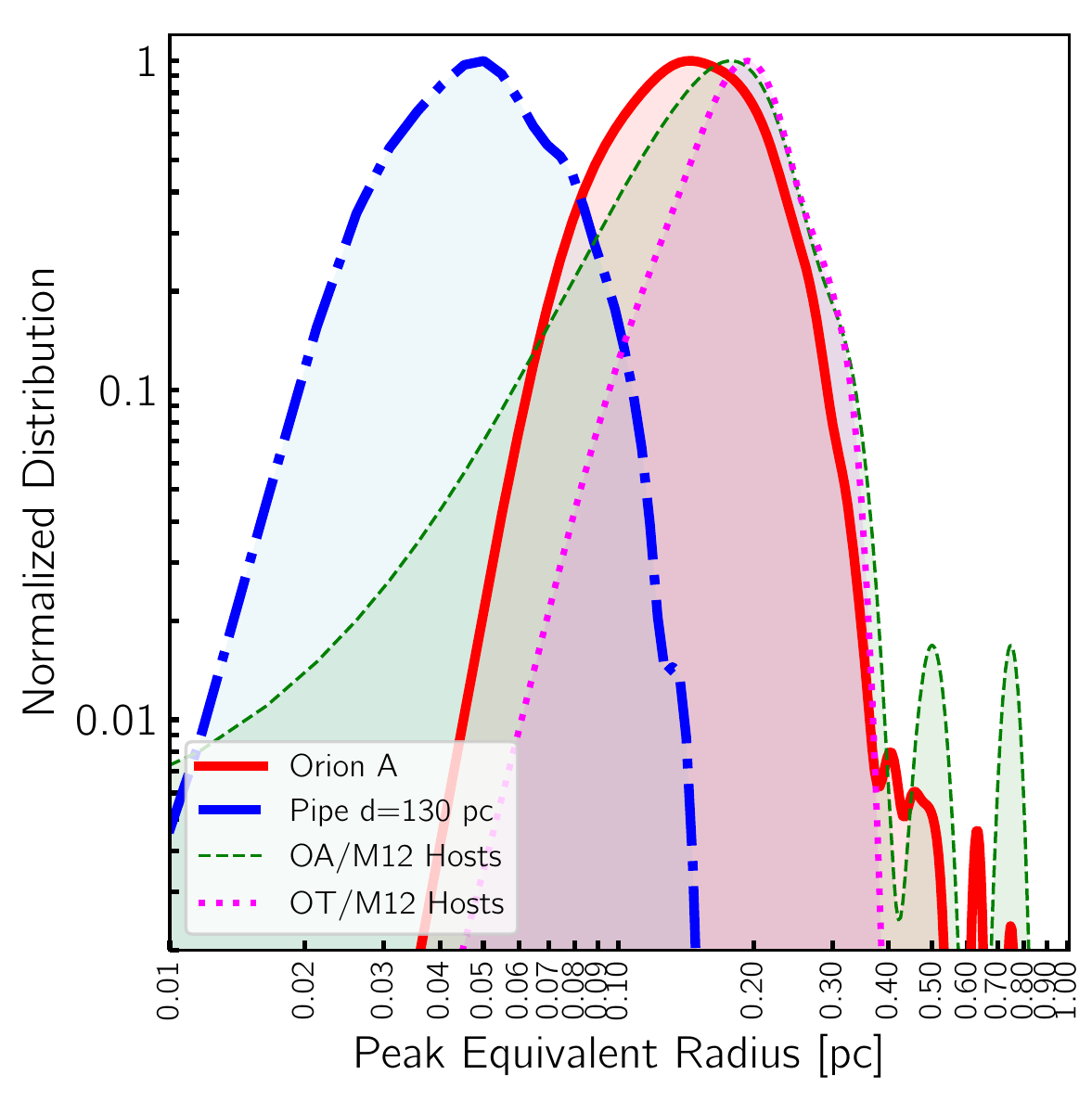}
\caption{Distribution of equivalent radii for the Pipe and Orion A column density peaks. The colour palette and organization for the KDEs is the same as for Figure \ref{fig:TdustHist}. \label{fig:ReqHist}}
\end{figure*}

\begin{figure*}
\includegraphics[width=3.3in, angle=0]{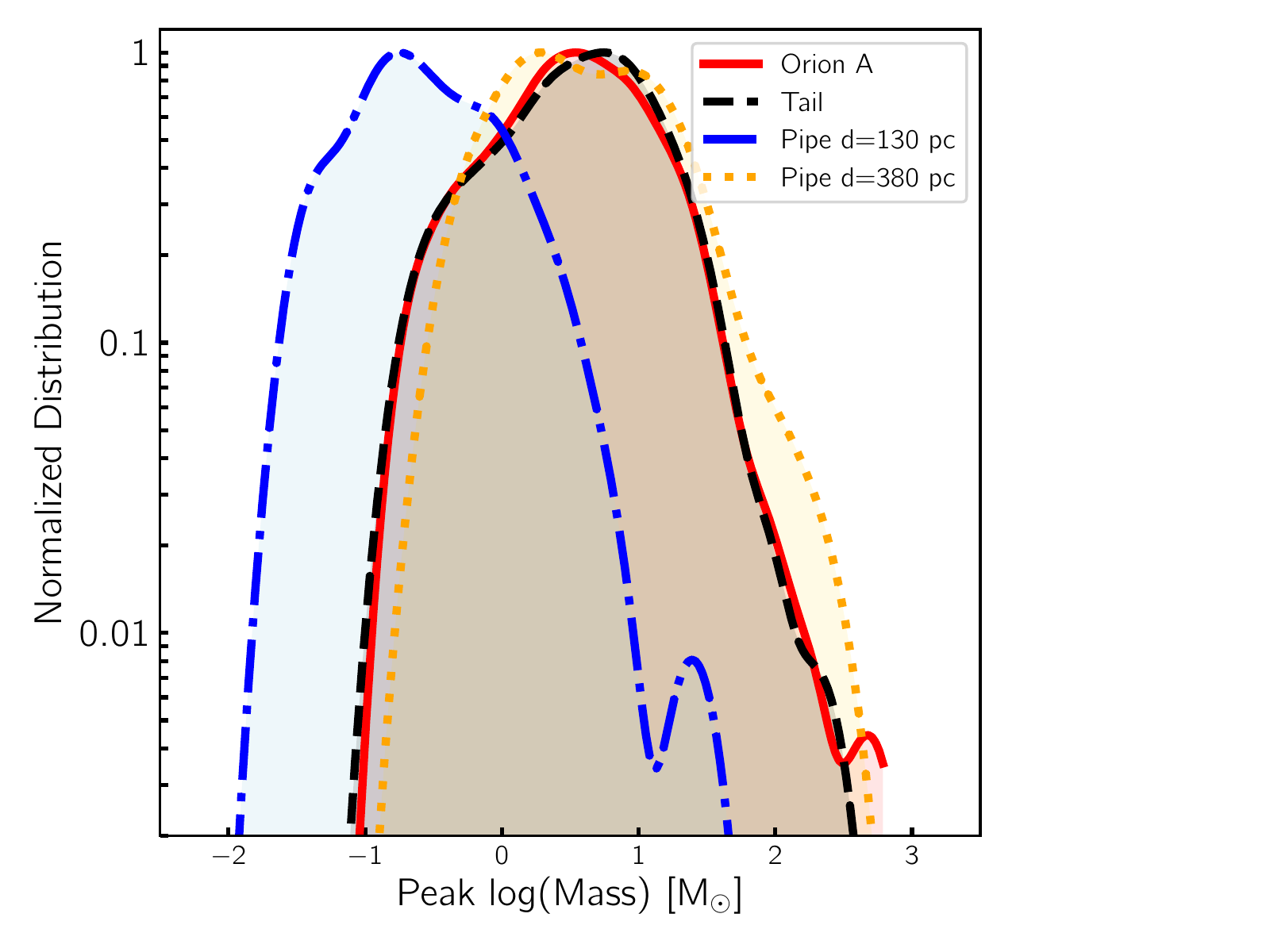}\includegraphics[width=3.35in, angle=0]{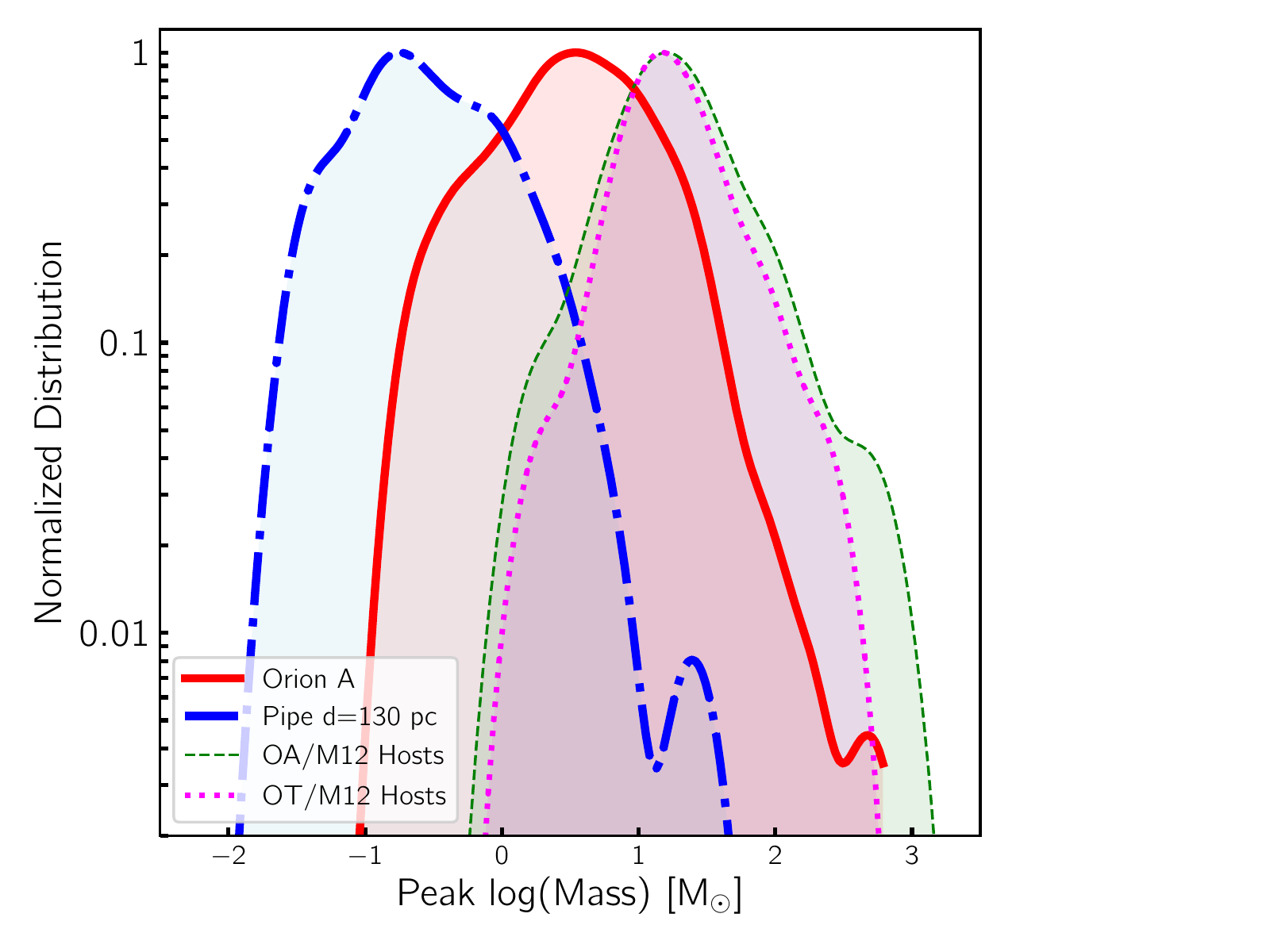}
\caption{Distribution of logarithmic mass for the Pipe and Orion A column density peaks. The colour palette for the KDEs is the same as for Figure \ref{fig:TdustHist}. \label{fig:MassHist}}
\end{figure*}

\begin{figure*}
\includegraphics[width=3.3in, angle=0]{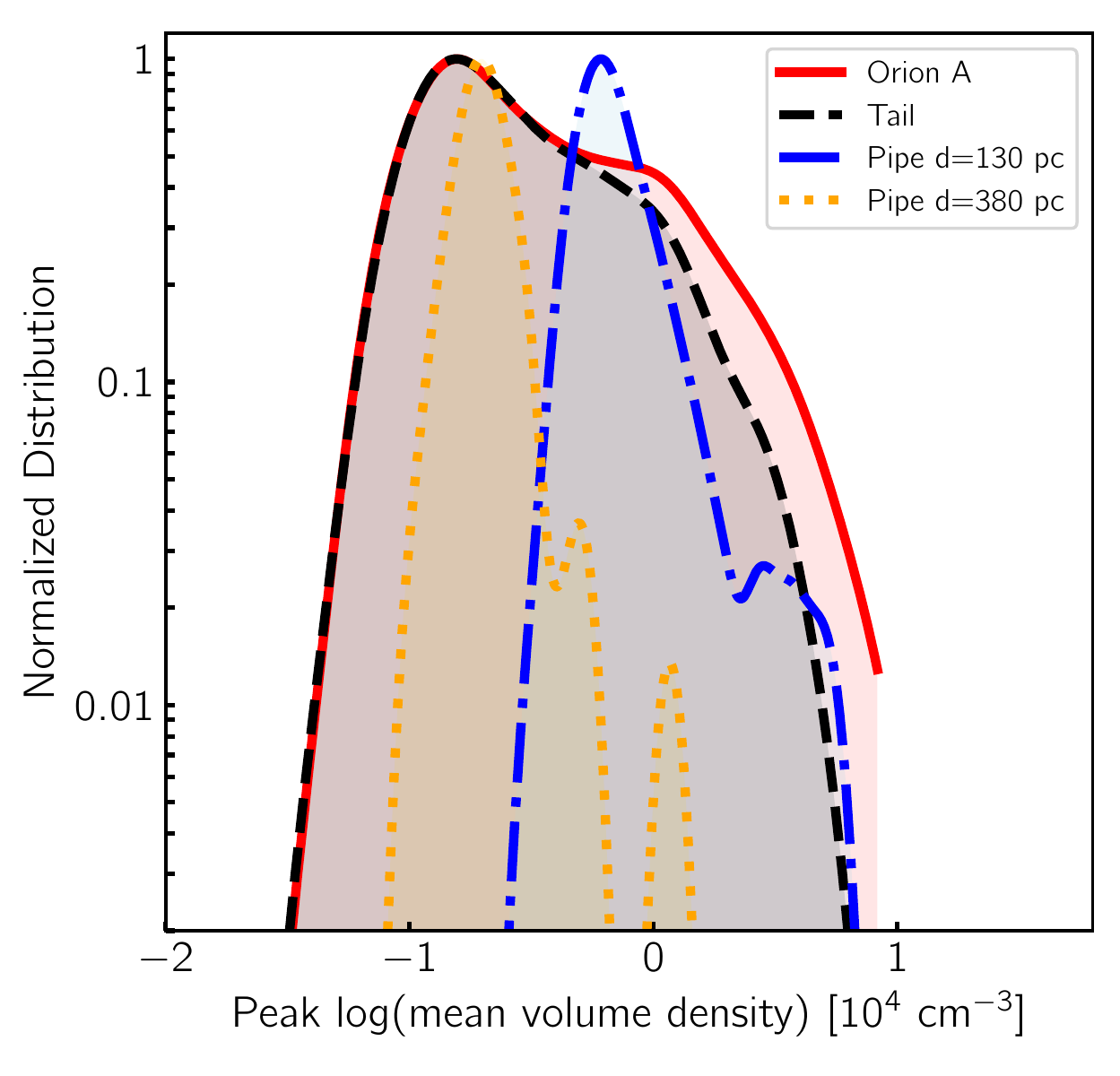}\includegraphics[width=3.3in, angle=0]{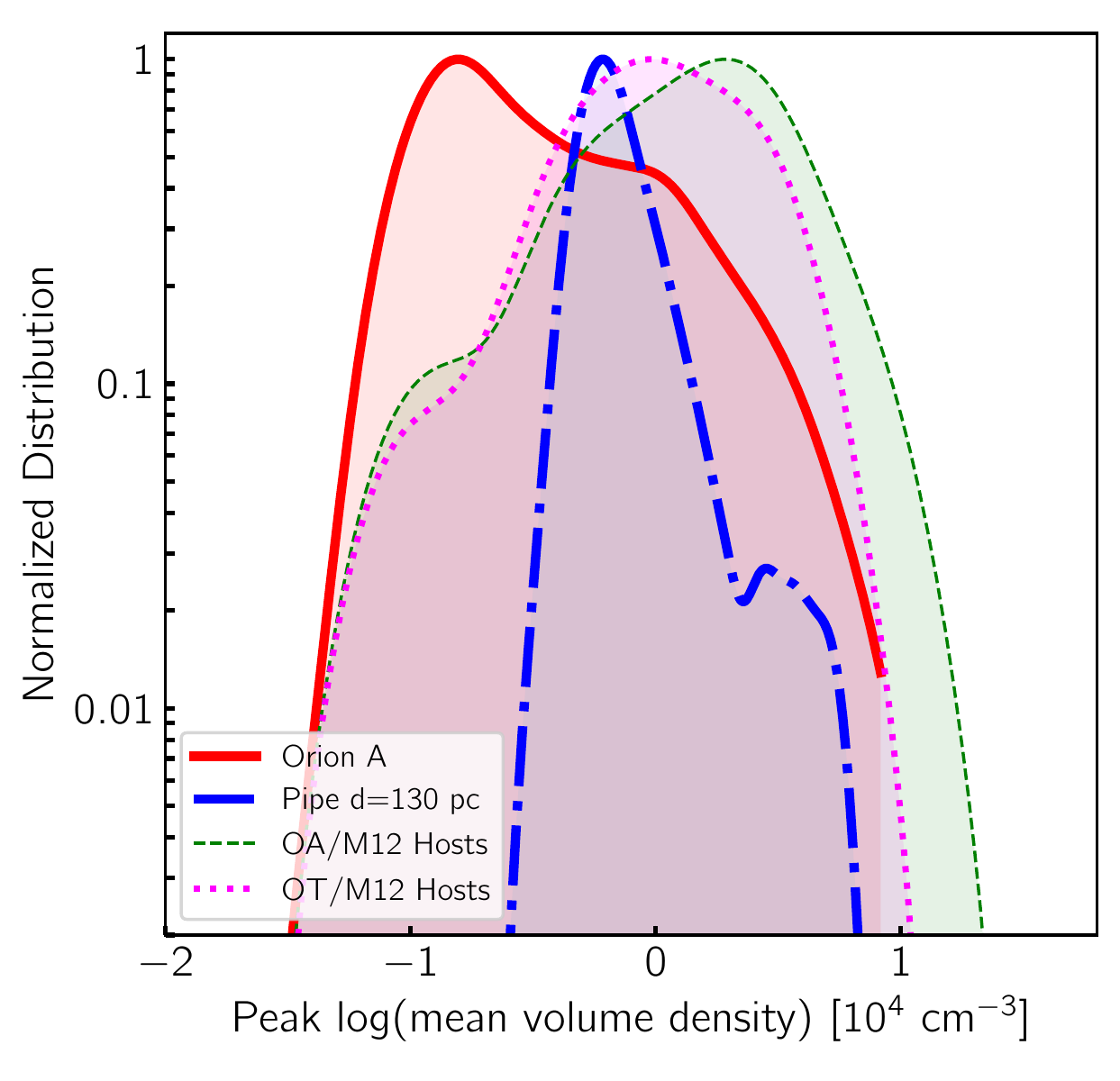}
\caption{Distribution of logarithmic volume density for the Pipe and Orion A column density peaks. The colour palette for the KDEs is the same as for Figure \ref{fig:MassHist} \label{fig:DensHist}}
\end{figure*}

We added two additional experiments: in the first experiment, we simulated how the Pipe would look if it were placed at the same distance as Orion A, by using the \texttt{Gauss} routine of the IRAF\footnote{IRAF is distributed by the National Optical Astronomy Observatory, which is operated by the Association of Universities for Research in Astronomy (AURA) under a cooperative agreement with the National Science Foundation.} package, to convolve the Pipe $A_V$ map with a Gaussian filter of 90$\arcsec$ FWHM (so the final angular resolution is 97''), but keeping the same pixel scale. We processed this Pipe convolved map (hereafter c-Pipe) as if it was a new one. The list of peaks for the convolved map is not just a scaled copy of the list for the full resolution Pipe map, because some peaks become smoothed, the background is also smoothed out and the regions of individual objects change. As a result of this, our method only yields 177 density peak identifications for the c-Pipe map. In the second experiment, we isolated a portion of the Orion A ``Whale Tail" region, by considering a subset of identified peaks in the region within $212<(l/\mathrm{deg})<215$, $-20<(b/\mathrm{deg})<18$. The Orion Tail contains a low star forming activity compared to the rest of the complex, and we considered it to provide a more fair direct comparison with the Pipe and c-Pipe maps. The Orion Tail area contains 489 column density peaks.

\section{Results }\label{s:results}

\subsection{Column density peak properties distributions }\label{s:results:ss:pdps}

In Figures \ref{fig:TdustHist}, \ref{fig:ReqHist}, \ref{fig:MassHist} and \ref{fig:DensHist}, we show the resultant distributions for average $T_d$, equivalent radii, mass and mean volume density after the peak identification with \texttt{CLUMPFIND-2D} in the Pipe, Orion A, c-Pipe and Orion Tail maps. 
The distributions are presented as Kernel Density Estimates (KDE), constructed with a Gaussian kernel truncated at 3-sigma. The KDE bandwidths were estimated with the Scott's rule \citep{scott79}. All KDEs shown in these figures were normalised independently to their maxima in each distribution. In the left panels, we compare the KDE of the four main maps (Orion A, Pipe, Orion Tail and c-Pipe), while on the right panels we show the distributions for peaks that contain YSOs from the list of \citet{megeath:2012aa} in the Orion A and Orion Tail clouds (``M12 hosts"), along with the Orion A and Pipe KDEs for comparison. We constructed the M12 hosts KDEs by cross-matching our peak positions with the YSO catalog of \citet{megeath:2012aa} using a tolerance radius of 1 arcmin.

From figure \ref{fig:TdustHist}, we see that the $T_d$ KDEs for the Orion A map indicate a relatively wide distribution that runs from about 11 K, peaks near 18 K and then fades down smoothly to a drop near 60 K. A second peak near 35 K may be indicative of the transition at the cluster forming region, where the emissivity changes significantly, as it can be seen in the map of Figure \ref{fig:OA_Av}. The Orion Tail values, are limited to the lower end of the distribution, (11 to \textbf{19} K with a peak at 16 K).  The Pipe and c-Pipe $T_d$ distributions, are also confined to a narrow range but 2 degree higher (13 to \textbf{21} K with a peak at 18 K). The peaks hosting YSOs from the list of \citeauthor{megeath:2012aa} are not limited to a particular range of the $T_d$ distribution in the Orion A map, but in the Orion Tail map, they are confined to a narrow range between 13 and 16 K. We also confirm a clear anti-correlation between $T_d$ and both the average and peak A$_V$ for the cores in all the three samples, as indicated in the study of \citet{birgit2018}.

The KDEs in figure \ref{fig:ReqHist} show that the projected size distributions of density peaks. The Pipe distribution is confined to sizes between 0.03 and 0.13 pc. This is consistent with the peak size obtained from a similar analysis of the near-infrared excess $A_V$ map in RAL10, which showed a distribution running from 0.04 to 0.15 pc. The distributions we obtained for the Orion A and the Orion Tail maps are very similar, with a lower limit near 0.08 pc.  The c-Pipe peak sizes are distributed in a range consistent with that of the Orion A and Orion Tail distributions. However, the c-Pipe KDE shows an excess of larger peaks respect to Orion above 0.3 pc and a tail towards values below 0.04 pc. The peaks hosting YSOs in Orion A do not appear to be constrained to any particular peak size, although the distribution is quite similar to that of the Pipe. The peaks hosting YSOs in the Tail between have a more constrained distribution, with a peak near 0.25 pc.

In Figure \ref{fig:MassHist} we show the peak mass distributions. Using our clumpfind method, the amount of mass extracted in the density peak identification was 216.1 and 3376 M$_\odot$ (about 3 and 5\% of the total) for the Pipe and Orion A, respectively. In the case of the Pipe map, the distribution is consistent with those obtained from low and high resolution extinction maps of the Pipe Nebula in \citet{rathborne:2009aa} and RLA10, with a lower end cutoff near 0.02 $\mathrm{M_\odot}$, and a maxium near 0.2 $\mathrm{M_\odot}$. The Orion A distribution shows a less narrow flat top with a maximum near 5 $\mathrm{M_\odot}$, which is consistent with the mass distribution from the Orion Tail map, showing a maximum near 4.5 $\mathrm{M_\odot}$. Both Orion distributions show a power-law like drop above 10 M$_\odot$. The YSO hosting peaks are distributed along the upper half of the distribution, with a peak near 30 $\mathrm{M_\odot}$. 

The c-Pipe distribution is, however, very consistent with those of the Orion and Orion Tail maps, showing also the power-law drop above 10 M$_\odot$. Instead of a maximum, the distribution has a flat top with a small dent that actually coincides with the Orion maxima near 5 $\mathrm{M_\odot}$. We applied a Kolmogorov-Smirnov test to the cumulative distributions of peak mass for the Orion A, Orion Tail and c-Pipe maps, and we obtained probabilities of 99.5 and 99.8 percent that the c-Pipe peak sample is obtained from a distribution identical to those from Orion and the Tail, respectively.
\begin{figure*}
\includegraphics[width=0.99\textwidth, angle=0]{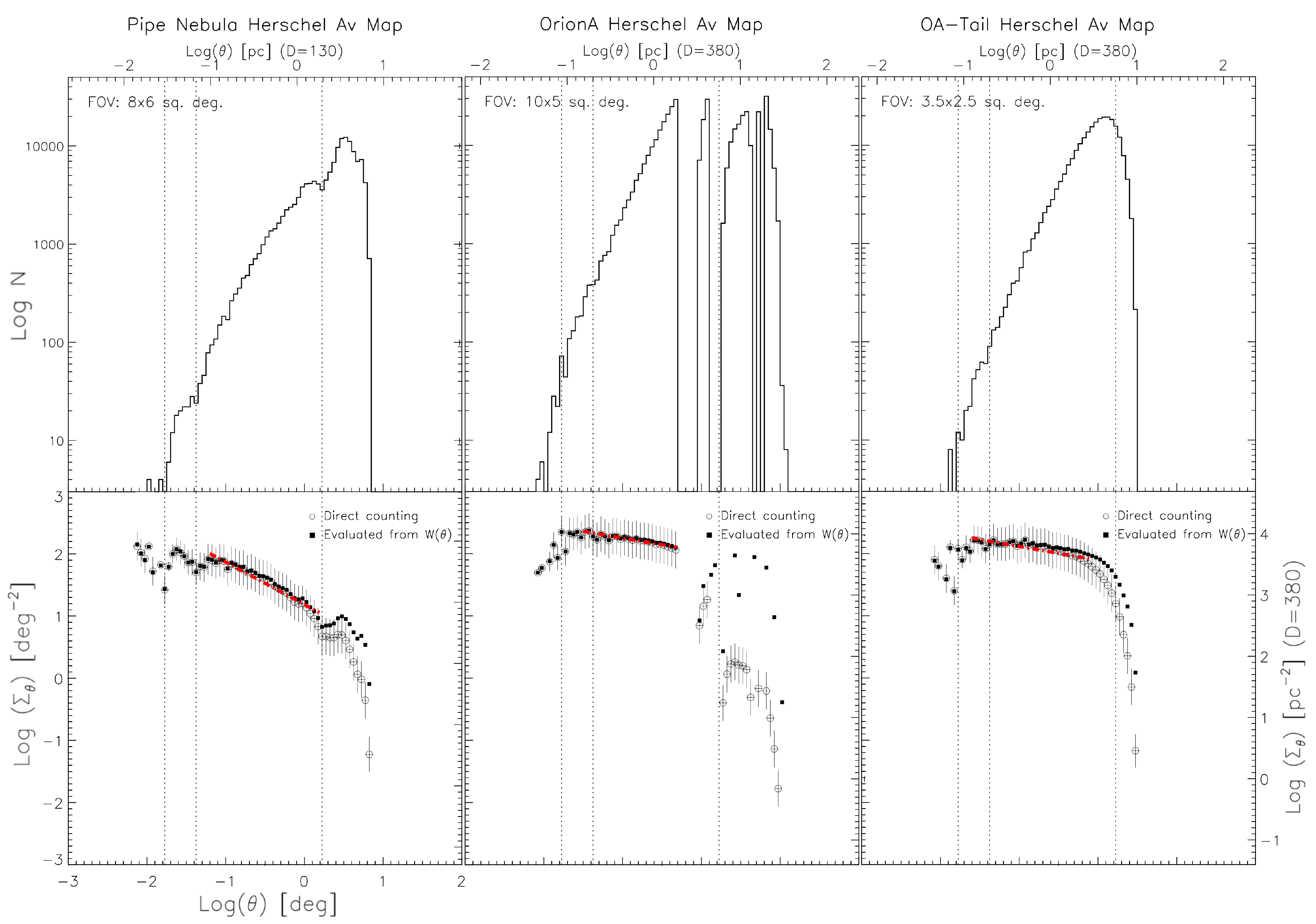}
\caption{Mean Surface Density of Companions Analysis for the Pipe, Orion A and Orion Tail HP2 maps (panels a, b and c, respectively). \textit{Top panels:} Distribution of angular separations between density peaks. The vertical dotted lines mark inflection points in the Pipe distribution at 0.04, 0.1 and 4 pc ($\log{\theta}\approx -1.8,-1.4\mathrm{\ and\ }0.2$), which also correspond to slcope breaks in the Pipe MSDC function below. The lines are shown at the same $\theta$ values in the Orion A and Orion Tail plots for comparison. \textit{Bottom panels:} MSDC as a function of angular separation. The white circle and black square symbols correspond to the $\Sigma_\theta$ estimates from ``direct counting'' and $W(\theta)$, respectively. The dotted red lines corresponds to power-law fits applied to the direct counting function in specific ranges (see text). \label{fig:MSDC}}
\end{figure*}

Last, we show the resultant mean volume density distributions in Figure \ref{fig:DensHist}. For the Pipe, the distribution is relative narrow, with a maximum at $0.7\times10^4\mathrm{cm^{-3}}$, a lower end drop near $2.5\times 10^3\mathrm{cm^{-3}}$ and an upper end drop near $3\times 10^4\mathrm{cm^{-3}}$.~In the case of the Orion A map, there is a maximum near $2\times10^3\mathrm{cm^{-3}}$ and an extended tail with pronounced drop above $10^4\mathrm{cm^{-3}}$. The Orion Tail catalog is distributed over a similar range, except that it shows a consistently smaller fraction of peaks with densities above $10^4\mathrm{cm^{-3}}$. The \citeauthor{megeath:2012aa} YSOs are clearly distributed along the most dense peaks. The c-Pipe map shows a distribution around much lower values than the Pipe at 130 pc because the convolution produces peaks with systematically larger sizes. However, the distribution is more consistent with those of Orion and the Tail, with a peak near $2\times10^3\mathrm{cm^{-3}}$; the distribution is still narrow as for the Pipe, with lower and upper end drops near 0.5 and 8$\times 10^3\mathrm{cm^{-3}}$. 

\subsection{Mean Surface Density of Companions }\label{s:results:ss:msdc}

The Mean Surface Density of Companions (MSDC) has proven to be a reliable tool for the exploration of the spatial distribution of stellar clusters or column density peaks in 2-dimensional maps. \citep[e.g.][]{gomez93,bate98,roman:2010aa,tafalla15,palau18,baobab19}. We applied this technique to our HP2 maps, providing a comparative scheme that allows us to explore the effects of a) clustering: in our case, the tendency in the spatial distribution of peaks from form pairs to form groups and sub-groups, and b) the scale regimes at which density peaks are distributed across the intrinsic structure of the cloud, in our case, how density peaks are distributed along filaments of different lengths; this is directly related to fragmentation. 

\begin{center}
\begin{figure*}


\includegraphics[height=5.0in, angle=0]{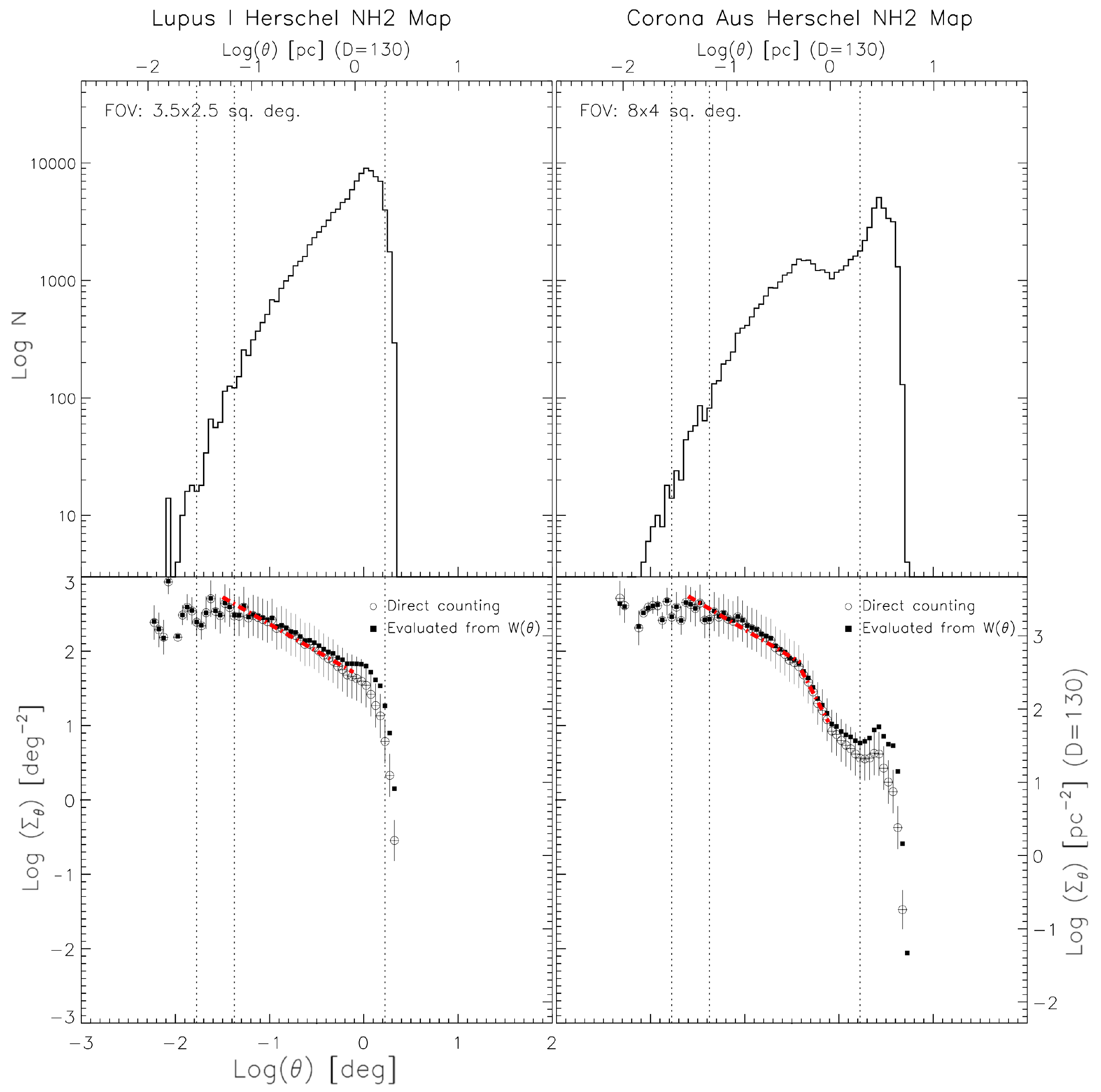}
\caption{Mean Surface Density of Companions Analysis applied to prestellar core lists extracted from N(H$_2$) maps of the Lupus I and the Corona Australis molecular clouds (panels a and b). The symbology for the plots is the same as in Figure \ref{fig:MSDC}. \label{fig:MSDC2}}
\end{figure*}
\end{center}

We used the prescription by \citet{simon97}, also applied by \citet{roman:2010aa,palau18} and \citet{baobab19} in studies of column density maps at different scale regimes. We started by determining the angular distances, $\theta$, from one to each other density peak in our lists, and organised the result in annular, logarithmic bins, 0.2 dex wide, within $-3.0<\log{\theta \mathrm{\ [deg]}}<2$. We calculated the MSDC, $\Sigma _\theta$ in two different ways: in one, we directly counted the number of elements in each annuli, $N_p (\theta)$ dividing by the area of the annuli and renormalising by the total number of elements. In the other, we used the Two-point Correlation Function (TPCF), in which $\Sigma _\theta$ is directly compared to a uniform, random distribution of points in a rectangular area, $A_{map}$, equal to that of the corresponding HP2 map. The TPCF is $W_\theta=(N_p/N_r)-1$, where $N_r$ was estimated from averaging counts in $10^4$ random fields simulated with a Monte Carlo routine. This way, we can compare directly the MSDC obtained from direct counts against $\Sigma _\theta=(N_p/A_{map})(1+W_\theta)$. Therefore, we can trust those points in the function where both estimates are the same within the uncertainty. For those points where both estimates diverge, the function is dominated by edge effects due to the use of a finite size map. In Figure \ref{fig:MSDC} we show the resultant MSDC functions from the Pipe, Orion A and Orion Tail clouds, along with a histogram for the separations, $\theta$. Below, we discuss each cloud: 

In the left panel (Pipe map), the top histogram shows how $\log{\theta}$ presents a uniform behaviour within 0.08 and 1.6 pc, decreasing linearly within the uncertainty, although a slight curving of the data points can be noticed. The smoothness of the data point curve is consistent with the MSDC obtained from the NICER map of RAL10 (see their figure 12) and tells us how coherent the structure of the cloud is over almost 2 orders of magnitude.  From a simple fit to a linear function we obtain a slope of -0.68: a linear behaviour with a slope close to -1 is indicative of uniform fragmentation along the filamentary structure of the cloud across the scales defined by the range. However, the slight curving of the function, with a small ``bump'', visible in the top histogram near 1.5 pc., is indicative of aggregation in small groups, i.e. clustering. The ``bump'' is actually consistent with RAL10, where a slight curving of the function is also detected, and where peaks were found to form groupings with separations narrowly distributed around $\sim 1.9$\ pc. For separations within 0.04 and 0.08 pc, the trend is also linear (although we can only resolve four bins) which possibly indicates a second regime of uniform fragmentation. The agreement between the MSDC estimated from direct counts and the TPCF reaches a lower limit near 0.015 pc (roughly 3000 AU), indicating the limit at which we resolve structure in the Pipe HP2 map.

\begin{figure*}
\includegraphics[width=6.8in]{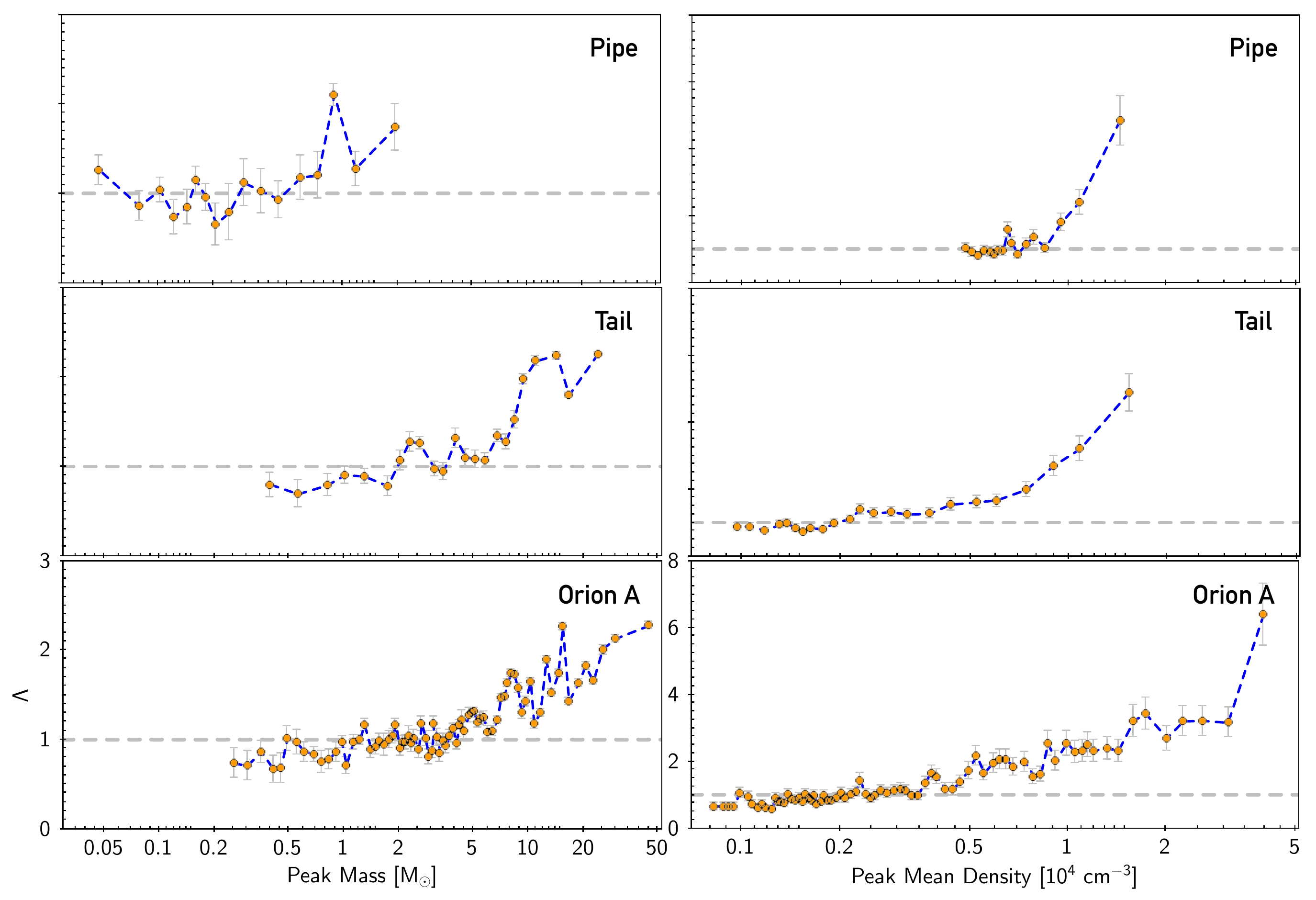}
\caption{MST $\Lambda$ parameter values as a function of peak mass for the Pipe (top), Orion Tail (middle) and Orion A (bottom) HP2 maps. Notice that the length of the axes is slightly different for some panels. \label{fig:MSTLambda}}
\end{figure*}

\begin{figure*}
\includegraphics[width=6.8in]{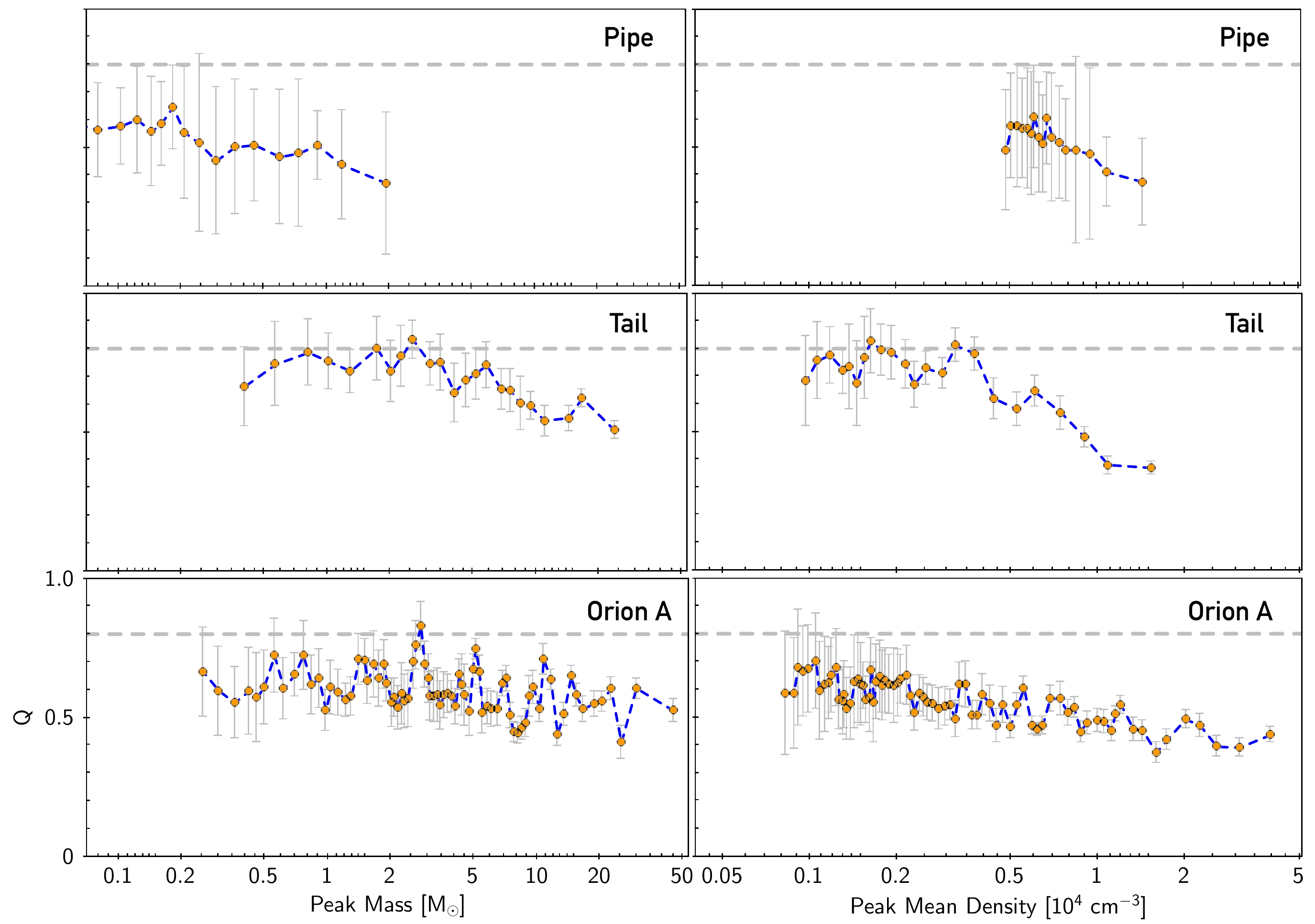}
\caption{MST $Q$ parameter values as a function of peak mass for the Pipe (top), Orion Tail (middle) and Orion A (bottom) HP2 maps. Notice that the length of the axes is slightly different for some panels. \label{fig:MSTQ}}
\end{figure*}

In the central panel (Orion A map) the histogram in the top shows a smooth and linear behaviour between $\sim 0.15$ and $\sim 1.8$ pc --a  similar regime as the Pipe--, although the slope for the Orion A MSDC (-0.22) is significantly more shallow than for the Pipe with no clear curving or "bumps". The smaller slope could indicate a less defined filamentary structure for the Orion A cloud. The absence of curving indicates that clustering of density peaks into small groups (substructure) is less obvious in Orion A compared to the Pipe. The linear regime is followed by an abrupt break near the 5 pc scale, where both the histogram and the MSDC function stop being coherent. This indicates the scale at which the cloud is disrupted by feedback from the stars of the Orion Nebula Cluster. At the lower end, the points within 0.04 and 0.15 pc, below the smooth regime described before, show a quasi-linear comportment but a positive slope, opposite to the Pipe.

In the right side panel we show the results for the Orion Tail map, which avoids the western side of the cloud. The histogram on the top shows a uniform quasi-linear behaviour between $\sim 0.1$ and $\sim 2$ pc, with a shallow slope (-0.25), similar to the one obtained for the Orion A map. However, instead of the disruption of the MSDC near 10 pc we see in the Orion A catalog, the function has a dramatic but smooth drop. There is an acceptable agreement of the two MSDC estimations (direct counts and $W(\theta)$) up to the 10 pc scale, indicating that the curving trend is a physical feature. Also, it is possible to notice a little more "bumpy" comportment of the data points. This indicates that uniform fragmentation, and possibly sub-clustering, are common features from scales near 0.1 pc --possibly sub-Jeans\footnote{For a cloud with a temperature of 10 K and an average density of $10^4\mathrm{cm^{-3}}$, the Jeans lenght is 0.2 pc, and hence 0.1 pc would be considered a sub-Jeans scale}--,  all the way to the scales of molecular clumps (star group forming structures).

For comparative purposes, we applied the same MSDC method to two additional list of pre-stellar cores from recent literature. We used the Corona Australis (CrAs) Molecular Cloud pre-stellar core catalog from \citet{bresnahan:2018} and the Lupus I core list  from \citet{benedettini:2015aa}.  Both included pre and proto-stellar structures. The distance and extension of these clouds are very similar to those for the Pipe Nebula, which make them ideal for a direct comparison while providing an adequate, independent check of the method.  In both cases the cores were identified and extracted from a N(H$_2$) map, not converted to visual extinction, but the analysis is equivalent, as our main concern is the spatial distribution and as we expect that extinction and column density are related linearly. The core extraction technique in both papers is based on the \texttt{getsources} code \citep{getsources:2013aa}. We assumed a distance of 130 pc for both clouds. The MSDC for Lupus I shows a well defined linear regime (a slope of -0.72) within $0.04\leq \theta<1.0$ pc, with a soft ``bumpy" trend, indicating a slight tendency for sub-clustering. The linear behaviour appears to show a relatively flat tendency within $0.02<\theta<0.04$ pc, below which it is disrupted down to the detection limit of the map, near 0.01 pc.  The MSDC for Corona Australis shows a much more obvious bump, which we fit as a double linear regime with two distinct slopes, -0.75 and -2.4, within $0.04<\theta\leq0.5$ and $0.5<\theta<1.0$ pc respectively. We think this is showing the division of the cloud into two regimes of uniform fragmentation (see figures 4 and 5 of \citeauthor{bresnahan:2018}), one in the regions CrA-A to C ($\mathrm{18^h:58^m}<\mathrm{RA}<\mathrm{19^h:05^m}$), where cores are more crowded along shorter filaments, and another for the long filament regions CrA-D to H ($\mathrm{19^h:05^m}<\mathrm{RA}<\mathrm{19^h:24^m}$). It is worth noticing that the flattening tendency below 0.04 pc is also present in the Corona Australis plot. The slopes for the linear regime in Lupus I and the first linear regime in Corona Australis are very similar to the one we obtained for the Pipe Nebula.  

\subsection{Minimum Spanning Tree Analysis }\label{s:results:ss:mst}

The spatial distribution of density peaks in the Pipe, and its relation to the distribution of mass and density was studied by AR18, who used the Minimum Spanning Tree (hereafter MST) edge length methodology \citep{allison:2010aa,MyC2011} on the density peak catalog of the Pipe extinction map of RAL10. Here, we applied a similar methodology to our three HP2 main maps (Pipe, Orion A, OATail), allowing for a comparative analysis.

We summarise our MST analysis as follows: The original idea behind this method is to investigate mass segregation in clusters. If the most massive stars are located near the center of a cluster, then their average projected distances to their closest neighbours should be systematically shorter. We know that once clusters are formed they rapidly evolve and, for compact clusters (half-mass radius of 0.3 pc or less) ejection of massive members can be significant, so it makes more sense to understand segregation at formation or early evolution stages. To quantify this effect at the level of molecular cloud cores, we use a peak catalog sorted down on a given parameter (in our case, mass or mean volume density) and organised in equal number bins, which overlap by a fixed number of common members, $s$. For each bin, containing $n_p$ peaks, we obtain a central value for their edge length distribution, $l$, which we compare to the value $l_r$ obtained for a random sample of the same size in the same interval. The experiment is repeated extensively ($10^4$ repetitions in our case) and the median ratio $\Lambda=(l/l_r)$ is calculated. For those bins where $\Lambda>1$ the interval is suspected to show a segregation. Also following AR18, we used the MST derived $Q$ spatial parameter \citep{CyW:2004}, to test for centrally condensed versus sub-structured distribution in the peak bins, using again mass and mean volume density as the tracing parameters. The $Q$ parameter is defined in terms of the mean angular separation, $\bar{\theta}$ and the mean edge length, $\bar{l}$, simply as $Q=\bar{l}/\bar{\theta}$. At each bin in the sorted peak distribution, values $Q\lesssim 0.8$ are indicate a spatial distribution more consistent with sub-clusters, i.e. multiple small groups with a less defined center. Values $Q>0.8$ are, instead, more consistent with a centrally condensed distribution. Uncertainties for $\Lambda$ and $Q$ were calculated for each mass and density bin using jackknife resampling \citep[see][]{efron:1981}. 

In the left side panels of Figure \ref{fig:MSTQ} we show the $Q$ parameter as a function of mass for the three HP2 maps. In the case of the Pipe, the $Q$ values decrease slowly as a function of mass and stay below 0.7 along the whole distribution, except for a slight increase near the maximum of the mass distribution near 0.2 M$_\odot$; this suggest that the spatial distribution of density peaks in the Pipe is consistent with marked sub-clustering around the more massive peaks. For the Orion A distribution, the $Q$ parameter values oscillate irregularly around 0.6 with a sharp maximum rising above 0.8 near 2.5 M$_\odot$; this indicates that the density peaks across Orion A agree better with a spatial distribution where they aggregate around several centres. Finally, for the Orion Tail the $Q$ values stay above 0.6 for all bins up to 10 M$_\odot$, after which it moves below, slightly. The function also shows a --slower-- rise to 0.8 near 2.5 M$_\odot$; this behaviour is consistent with that for the whole Orion A map. This rather suggests that less evolved regions in the complex of Orion A present a spatial distribution similar to that of the entire complex, and so the spatial distribution in the gas still reflects early stages of the cloud evolution.

In the right side panels of Figure \ref{fig:MSTQ} we show the $Q$ parameter as a function of mean volume density for the three HP2 maps. In the case of the Pipe, the values stay below or close to 0.6 for the whole distribution. Just as in the case for mass, the uncertainties are large, but the comportment of the function is relatively smooth, decreasing monotonically as a function of density. This indicates that the amount of substructure also increases around the most dense cores in the Pipe. The decreasing trend for $Q$ is also present in the Orion A map, except that the function is less smooth, and for the low density bins, $Q$ oscillates near 0.7, indicating that the less dense bins pile up around fewer centres. Notice also, that the density range for the Orion A is significantly wider than the Pipe, but in the 0.5-1.5$\times 10^4\mathrm{\ cm^{-3}}$ interval, the Orion A appears to have a similar level of substructure as the Pipe. In the bottom plot of the figure, we see how $Q$ is closer to 0.8 below $4\times 10^4\mathrm{\ cm^{-3}}$, suggesting that the Orion A Tail gas has a less marked tendency to break into smaller groups around higher density centres. This is consistent with the shallow slope we observe in the MSDC for the regime of small angular separations.

In the left side panels of Figure \ref{fig:MSTLambda} we show the $\Lambda$ parameter as a function of mass for the three HP2 maps. In the case of the Pipe, we see how the parameter oscillates near 1 until around 1 M$_\odot$, when it suddenly rises above 1.5 for the last few bins. This is indicative of a light mass segregation effect near the most massive objects in the Pipe, which is consistent with the result on the extinction map in AR18. Moving onto the Orion A map, we now see a more clear, steady rising tendency from low to high mass, which is maintained over almost two and a half orders of magnitude, indicating that mass segregation is a more clear effect in a cloud like Orion A. We can confirm this behaviour in the Orion Tail map. 

\begin{figure*}
\includegraphics[width=6.0in]{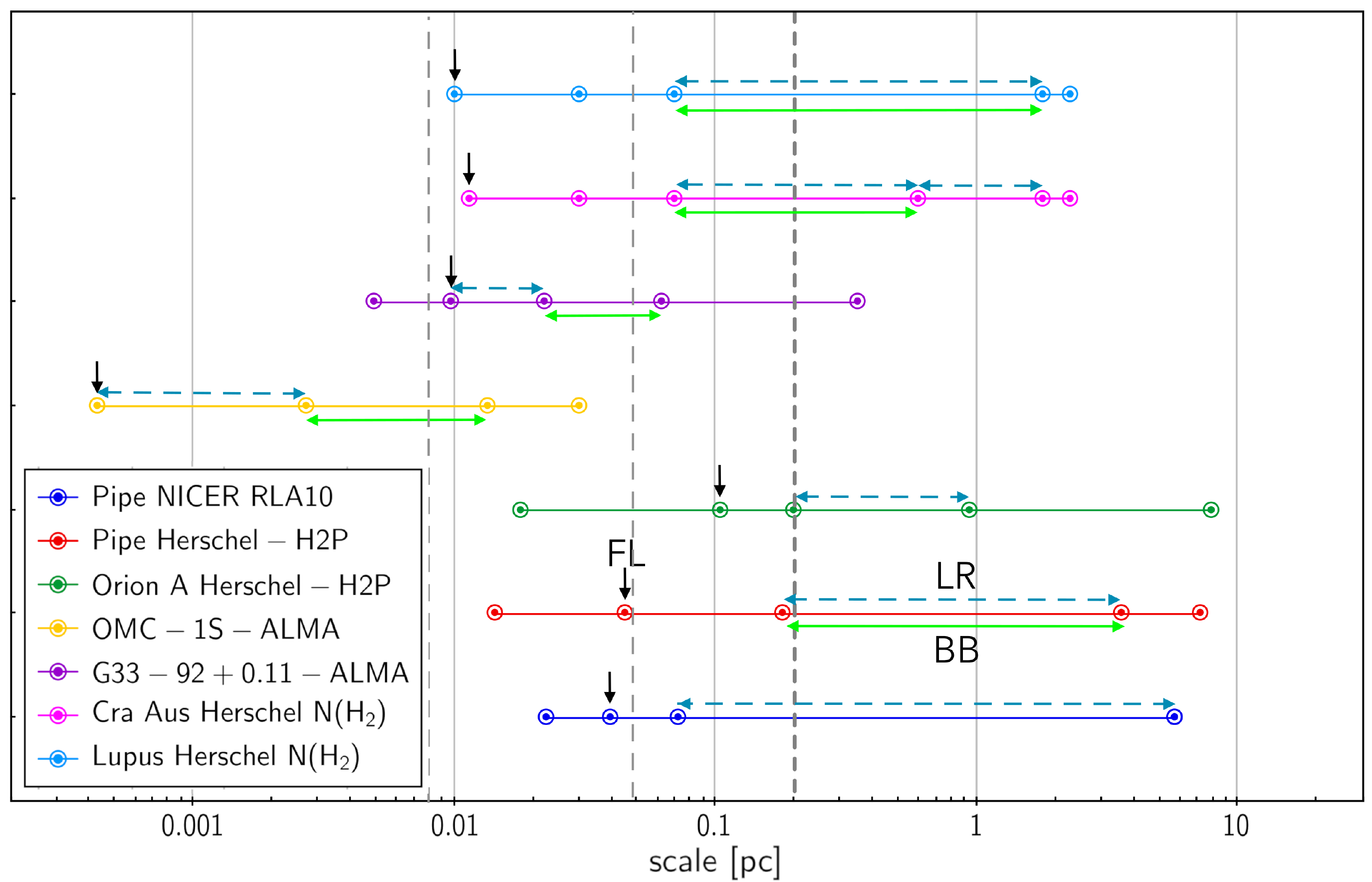}
\caption{Scale regimes for the MSDC in distinct GMC region maps, each identified by a cooler. Each graph shows the map limits as extremes, with possible fragmentation scale limits (FL) indicated by the small black, vertical arrows. The dashed, blue double-arrow tip lines indicate a linear behaviour (LR), while the green, solid double-arrow tip lines indicate regimes where the MSDC is "bumpy" (BB) indicating significant presence of sub-cluster structure. The vertical, thick dotted line shows the Jeans length for a cloud with a temperature of 10 K, and a mean density of $10^3\mathrm{\ cm^{-3}}$. The two vertical thin dashed lines mark the mean Jeans length derived for higher density regimes in the studies of \citep{palau18} and \citet{baobab19}.\label{fig:scales}}
\end{figure*}

In the right side panels of Figure \ref{fig:MSTLambda} we show the $\Lambda$ parameter as a function of mean volume density for the three HP2 maps. For the Pipe map, the parameter fluctuates around a value of 1 for the low density bins, but close to $0.8\times 10^4\mathrm{\ cm^{-3}}$ the values rise quickly all the way to $\Lambda=5$ in the last bin. This confirms the density segregation effect observed by AR18 in the Pipe extinction map, except that it is more blunt in our HP2 map. For the Orion A map, the steady increasing tendency of $\Lambda$ as a function of mass, is also present as a function of density all the way to $5\times 10^4\mathrm{\ cm^{-3}}$, surpassing $\Lambda=2$ in the bins above $0.5\times 10^4\mathrm{\ cm^{-3}}$, showing a very sharp peak in the last bin. For the Orion Tail map, where the main cluster regions are removed, it also increases as a function of density, except that the strongest segregation effect is noticeable from $1\times 10^4\mathrm{\ cm^{-3}}$, similar to the Pipe.

We also applied the MST analysis to the c-Pipe map and confirmed that the bearing of the $Q$ and  $\Lambda$ parameters are conserved (evidently, with less resolution). The corresponding Figure is in Appendix \ref{app2}.

\section{Discussion and Summary}\label{s:discussion}

In this work we presented a comparative study of the physical properties and spatial distribution of density peaks in two giant molecular clouds: the Pipe Nebula, with minimal star formation, and Orion A, with active cluster formation. Below, we discuss some of the similarities and differences we found, in the context of the origin of star clusters: 

\subsection{Strong similarities between Orion A and the Pipe Nebula \label{s:discussion:ss:similarities}}

Using the c-Pipe map that emulates a change in distance, we see that the peak size ($R_{eq}$) distribution shares the same peak as Orion A, but it also results in a wider distribution of sizes, with an extended tail towards low values and with an excess with respect to Orion for objects larger than 0.3 pc. Possibly, we do not observe peaks this large in Orion because those correspond to the scales where star cluster formation has already ocurred and gas is no longer concentrated in a clump. 

The peak mass distributions for the c-Pipe, Orion A and Orion Tail have similar ranges, sharing a peak near 5 M$_\odot$, and a similar power law drop above 10 M$_\odot$. This is an important feature that suggests that the conversion of gas into stars is a process that may be independent of the dimensions of the cloud. The fact that the mass distributions for density peaks in the c-Pipe and the Orion A maps are so similar is consistent with the idea that star formation is not determined by the actual amount of dense gas but on the material accretion process \citep[e.g.][]{burkert17}: the mass in dense cores in the two clouds may be very different, but the resultant distribution of dense peaks scales-up consistently, which indicates homologous fragmentation processes.

In terms of volume density, the systematically smaller objects of the Pipe results in a density distribution similar to the higher end for the Orion A and OATail, but once we convolve the Pipe map, we obtain that the Pipe compares well with the less dense, more quiescent regions of Orion, which contain only a minority of the star forming activity. In terms of the spatial distribution, the MSDC functions shown in Figure \ref{fig:MSDC} indicate that both Orion A and the Pipe have a regime where groups of density peaks distribute uniformly along filaments of different scales. This suggests again that, in both clouds, the fragmentation processes leading to the formation of clusters is equivalent, and only reflects different stages of cluster evolution. The MST analysis for both clouds show strong evidence that peaks are strongly segregated by volume density and a less clear segregation by mass. This confirms the results of AR18 and indicates that density segregation could be a common feature in star forming regions.

\subsection{Differences between Orion A and the Pipe Nebula }\label{s:discussion:ss:differences}


The MSDC function morphology indicates a wider regime of uniform distribution along filaments for the Pipe. Instead, in Orion A, such a uniform regime, and the aggregation in small groups are both less conspicuous. This suggests that the cluster precursor sub-structure we clearly observe in the Pipe, may be diluted in the more evolved Orion A cloud, as cluster structure is transferred from the gas into the stars. However, we can compare our Pipe MSDC function with the one obtained for the OMC-1S region which explores fragmentation and sub-cluster structure at a much smaller scale near 1000 AU using ALMA  \cite{palau18}, and confirms a MSDC behaviour very similar to the Pipe at small scales. Our MST analysis also shows that density peak sub-clustering permeates in the Pipe, while it is already less persistent across the remaining molecular gas in the Orion A. We confirmed that Lupus I and Corona Australis show a similar MSDC behaviour (see below), and this implies that the aggregation of density peaks in small groups could be an early feature of pre-stellar material, and is less evident in a more evolved cloud like Orion A. Finally, it is interesting to notice that the Orion A cloud appears to reach a slightly lower limit in $T_d$ (13 vs 16 K) despite having much stronger stellar feedback. However, we have to take into account that the Pipe Nebula is located closer to the Galactic Plane than Orion A, with potentially more warmer structures along the line of sight. Since the dust temperature is derived from a single modified-blackbody model, these warmer structures can bias the derived temperatures towards higher values. Still, we know that column density and temperature are anti-correlated and density peaks are typically associated with temperature minima \citep{birgit2018}. It may be interesting to consider if the star forming activity could permit more efficient cooling processes in an active cloud like Orion A.
 
 \subsection{Comparison to other star forming regions: the MSDC and thermal Jeans fragmentation }\label{s:discussion:ss:jeans}

The MSDC functions we obtained for Lupus I and Corona Australis have the same general behaviour as the Pipe. The plots of Figure \ref{fig:MSDC2} show regimes of uniform fragmentation across filaments in both clouds, in similar ranges. In the case of Lupus I, the curving of the MSDC is almost identical to the Pipe, and the apparent double slope in Corona Australis indicates that MSDC curving may be due to various regimes of fragmentation along the filamentary structures. \textit{Sub-cluster structure then, could be a direct consequence of the spatial distribution of cores across filaments}. 

In Figure \ref{fig:scales} we show a comparison of the scales involved in different MSDC functions derived with our methodology in the present and recent papers. The diagram illustrates the extent of the spatial scale regimes where a linear fit can be applied, as well as those regimes where a curve or bump can be observed in the MSDC. For the Pipe, Lupus I and Corona Australis, the linear and the bumpy regimes coincide well, which is not surprising given the physical similarities among these three clouds.

For the two regions observed ALMA, OMC-1S \citep{palau18} and G33+1.92 \citep{baobab19}\footnote{Notice that in both of these datasets, large scale emission is filtered out}, the linear behaviour is observed in a more narrow scale regime: The OMC-1S map explores a significantly smaller scale regime, resolving structures down to a few tens of AU, and the G33.92+0.11 map corresponds to a distant massive star forming region. but the resolution compatible with our Herschel (HP2 and $\mathrm{N(H_2)}$) maps of the Pipe and Orion A. The areas of the ALMA fields are much smaller, but the superior resolution ALMA provides much more detail than Herschel about the process by which thermal Jeans fragmentation across individual filaments \citep{palau14,palau15,palau18,pokhrel18} translates into the cluster organisation we observe across extended areas. In both the OMC-1S and G33.92+0.11 studies, the MSDC aid the authors to discuss how the observed spatial distribution of cores is compatible with thermal Jeans fragmentation. In RAL10 and AR18 they also proposed Jeans fragmentation as the main driver behind the spatial distribution of density peaks in the Pipe.

The MSDC indicates that the Pipe Nebula is currently more sub-clustered than the Orion Tail, which in turn is more sub-clustered than Orion A. Lupus and CrA are also sub-clustered and also have low star formation rates. This indicates that once clouds reach a high star formation rate like Orion A, their parental gas sub-structure we still observe in clouds like the Pipe, Lupus and CrA is diluted. The molecular cloud sub-structure in a cloud like Orion A was probably superceded by the formation of a large star cluster like the ONC.

\subsection{On the origin of the Schmidt conjecture }\label{s:discussion:ss:schmidt}

The study by \citet{lada:2013aa} showed that the empirical scaling relation of \citet{schmidt59} is well determined within clouds like the Pipe and Orion A, as well as in others like Orion B, Taurus and California. Schmidt's conjecture states that the star formation activity is directly related to the density structure of the cloud. \citeauthor{lada:2013aa} also discussed how star formation activity appears to be enhanced in regions where clouds surpass a certain column density threshold. \citet{burkert13} discussed how the Schmidt-type behaviour in molecular clouds indicates how the cloud forms dense gas across a continuum of scales. 

We think that the Schmidt relation could be imprinted in the volume density segregation of the molecular gas cores, described in AR18 and now confirmed in our MST analysis. Segregation by density (and to a lesser degree by mass) could be key to understand how clouds organize to form clusters. For instance, as recently shown by \citet{pavlik19}, the observed properties of the ONC are more compatible with a primordially mass segregated cluster model but tend to disagree with a non-segregated early configuration. Also, in a recent study, \citet{dib:2018aa} performed MST analysis on a different set of clouds, and showed a mild correlation between the segregation parameter, $\Lambda$, and the level of star formation activity. They suggest that this relation could be established at the earliest phases of evolution.

A cloud like the Pipe is the closest case we know of a delta-function event cloud, where only one of its cores has formed stars. The Pipe may be still evolving towards a higher level of star formation activity but, as shown by ARL18 and the present study, the molecular gas peaks show very strong segregation by density and a mild segregation by mass. Also, the Orion Tail shows stronger density segregation compared to the whole cloud that includes the formed and evolved cluster. The density segregation of molecular gas cores thus appears to be important at earlier stages of cloud evolution. If volume density acts as an early phase proxy for star formation activity, then the Pipe could be showing an early stage version of the relation found by \citet{dib:2018aa}.

In the scenario we propose, the spatial distribution of star clusters in more evolved clouds like Orion A is originated in the early spatial distribution of molecular cloud fragments. This is what we see in young clouds like the Pipe, with gas volume density appears to be the leading parameter for fragmentation and spatial distribution of pre-stellar structures.

\section*{Acknowledgements}

We want to thank an anonymous referee who made a throughful review of the original manuscript and provided useful comments that improved the quality and content of this paper. CGRZ and AP acknowledge support from program UNAM-DGAPA-PAPIIT Mexico, projects IN108117 and IN113119, respectively. EJA acknowledges support from the Spanish Government Ministerio de Ciencia, Innovacion y Universidades through grant AYA2016-75 931-C2-1 and from the State Agency for Research of the Spanish MCIU through the ``Center of Excellence Severo Ochoa" award for the Instituto de Astrofisica de Andalucia (SEV-2017-0709). In the present paper we analyzed data products based on observations performed with the ESA Herschel Space Observatory \citep{herschel0}, in particular employing Herschel's large telescope and powerful science payload to do photometry using the PACS \citep{herschel1} and SPIRE \citep{herschel2} instruments.
Based on observations obtained with Planck (http://www.esa.int/Planck), an ESA science mission with instruments and contributions directly funded by ESA Member States, NASA, and Canada. This publication makes use of data products from the Two Micron All Sky Survey, which is a joint project of the University of Massachusetts and the Infrared Processing and Analysis Center/California Institute of Technology, funded by the National Aeronautics and Space Administration and the National Science Foundation. We made use of TOPCAT \citep{topcat} and SAOimage DS9 astronomical software \citep{ds9}, as well as Python Matplotlib \citep{matplotlib}.








\appendix

\section{Wavelet Filtering}\label{app}

\begin{figure*}
\includegraphics[width=7in]{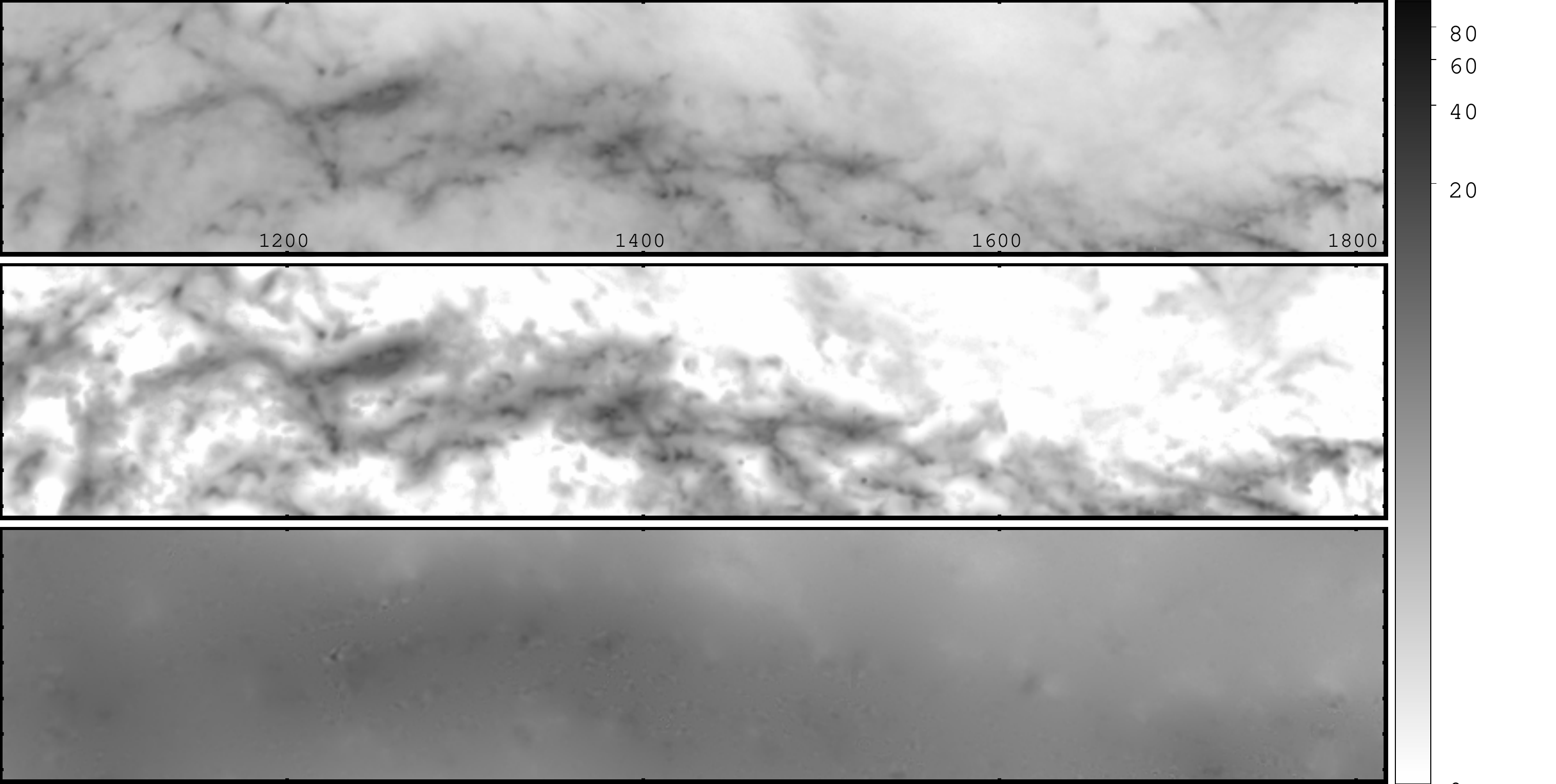}
\caption{The three panels show, from top to bottom, the same selected region in the HP2 image for the Orion A map in a scale of pixels. The top panel shows the original extinction map, the middle shows the image after processing with the MVM algorithm (wavelet filtering), and the bottom panel shows the residual after subtracting the second from the first image. It is clear that the wavelet processing helps to delineate and enhance the filamentary structure of the cloud where density peaks are located, and that the removed flux corresponds mostly to the low density background. The colorbar indicates visual extinction in units of magnitudes.  \label{fig:orionamvm}}
\end{figure*}

\begin{figure*}
\includegraphics[width=7in]{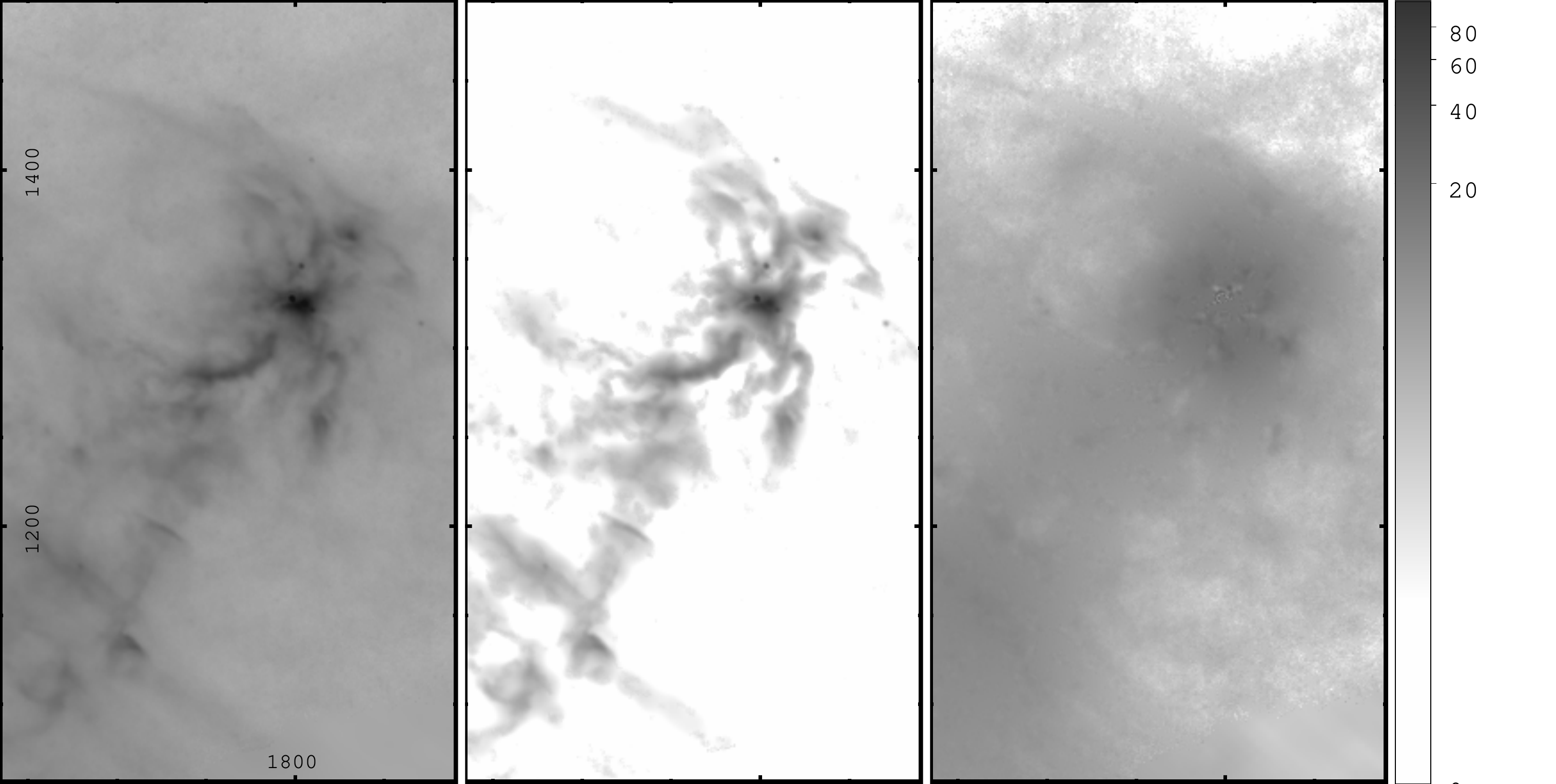}
\caption{Same as Figure \ref{fig:orionamvm} but for a selected region in the Pipe Nebula HP2 map. \label{fig:pipemvm}}
\end{figure*}

In figures  \ref{fig:orionamvm}  and  \ref{fig:pipemvm} we illustrate the process of wavelet filtering using the MVM method. The images show how the MVM algorithm systematically removes the low density background emission, enhancing the filaments and other structures where column density peaks are located in the clouds. 

\section{MST analysis for the c-Pipe map}\label{app2}

\begin{figure*}
\includegraphics[width=7.0in]{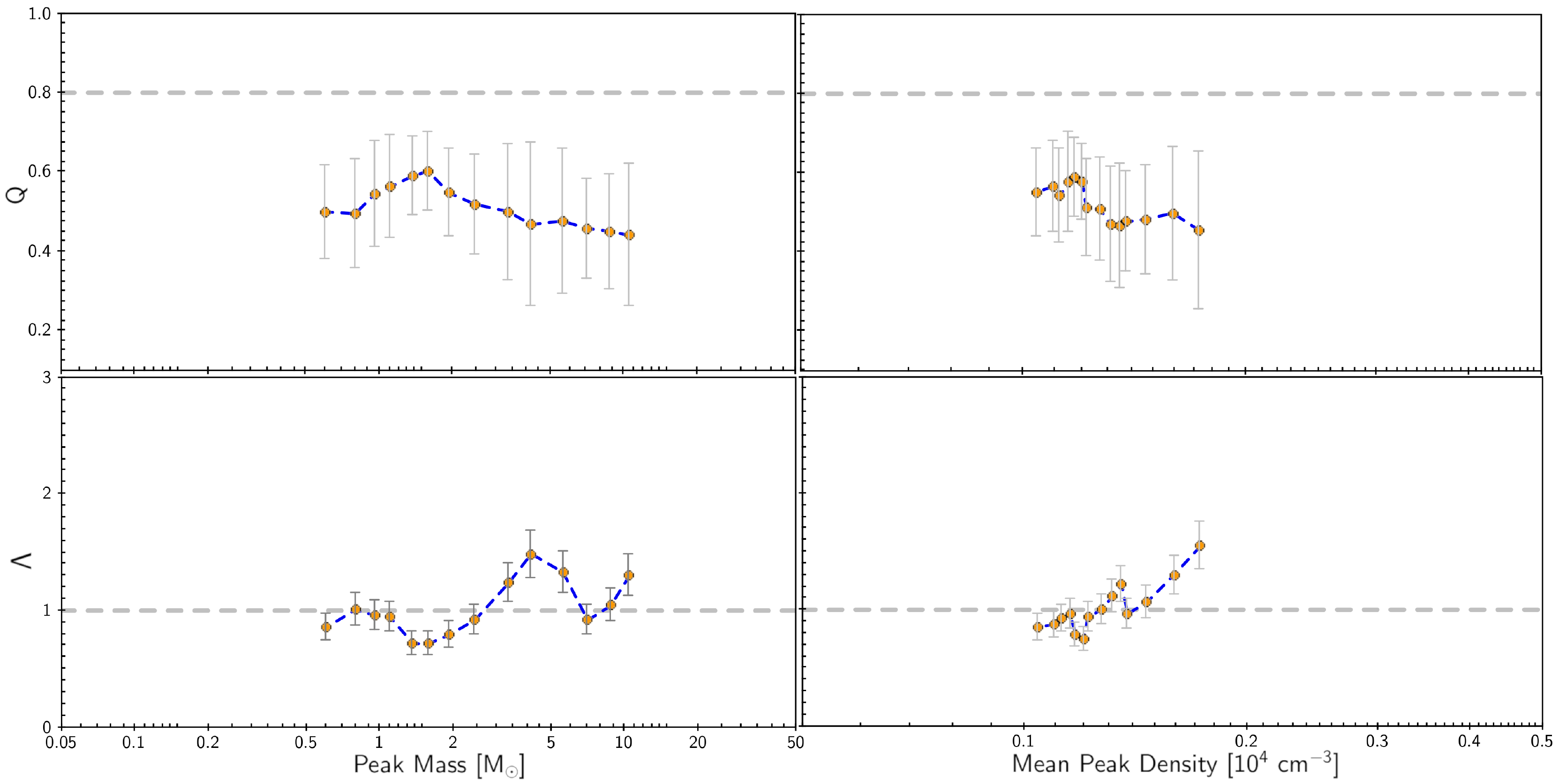}
\caption{MST $Q$ and  $\Lambda$  parameter values (top and bottom) as a function of peak mass and density (left and right) for the convolved Pipe Map (c-Pipe). \label{fig:MSTc-Pipe}}
\end{figure*}

We applied the MST analysis to the c-Pipe map and we present the corresponding plots in Figure \ref{fig:MSTc-Pipe}. The smoothing of the map imprints some blurring to the MST analysis as the number of peaks is reduced. However, the general behaviour observed for the full resolution Pipe map is kept. The $Q$ values stay consistently below 0.8 and fluctuate around 0.5, with a decreasing tendency as a function of both mass and density within the uncertainty. In the case of $\Lambda$, as a function of mass there is a fluctuation about 1.0, with a local maximum ($\Lambda \sim 1.5$)  near 4 M$_\odot$  and a second increment above 7 M$_\odot$. Thus, in general we may say that $\Lambda$ increases above the maximum of the peak mass distribution. As a function of density, the $\Lambda$ parameter also fluctuates around 1.0 until $0.4\mathrm{\ cm^{-3}}$, from where it increases consistently, but not as dramatically as for the full resolution counterpart. 

\bsp	
\label{lastpage}
\end{document}